\documentclass[twocolumn,showpacs,preprintnumbers,amsmath,amssymb,floatfix]{revtex4}
\usepackage{graphicx}    % Include figure files
\usepackage{dcolumn}    % Align table columns on decimal point
\usepackage{bm}              % bold math
\usepackage{amsmath,amsfonts,amssymb}
\usepackage{graphicx,epic,eepic}
\usepackage{subfigure}
\usepackage{enumitem}
\usepackage{graphics}
\usepackage{units}      %  \unitfrac[1]{C}{s}=\unit[1]{A}
\usepackage{mathrsfs}   % \mathscr{L} instead of \mathcal{L}
\newcommand{\preup}[1]{\,^{#1}\!}
\def\etal{{\emph{et al.}}}
\usepackage{verbatim}
\newcommand{\bd}[1]{{\bf{#1}}}
\begin{document}

\setlength{\pdfpageheight}{\paperheight}
\setlength{\pdfpagewidth}{\paperwidth}
%%%%%%%%%%%%%%%%%%%%%%%%%%%%%%%%%%%%%%%%%%%%%%%%%%%
\preprint{astro-ph/1000000}
%%%%%%%%%%%%%%%%%%%%%%%%%%%%%%%%%%%%%%%%%%%%%%%%%%%
%%%%%%%%%%%%%%%%%%%%%%%%%%%%%%%%%%%%%%%%%%%%%%%%%%%
\title{General Relativistic Magnetohydrodynamic Bondi--Hoyle Accretion}
%%%%%%%%%%%%%%%%%%%%%%%%%%%%%%%%%%%%%%%%%%%%%%%%%%%
%%%%%%%%%%%%%%%%%%%%%%%%%%%%%%%%%%%%%%%%%%%%%%%%%%%
\author{A.J.~Penner}
\email{a.penner@soton.ac.uk}
\affiliation{School of Mathematics, University of Southampton, Southampton SO17 1BJ, UK,\\ Department of Physics and Astronomy, University of British Columbia,
6224 Agricultural Road, Vancouver, B.C. V6T 1Z1 Canada}
%
%%%%%%%%%%%%%%%%%%%%%%%%%%%%%%%%%%%%%%%%%%%%%%%%%%%%
\date{November 9 2010; Revised XX; \LaTeX-ed \today}
%%%%%%%%%%%%%%%%%%%%%%%%%%%%%%%%%%%%%%%%%%%%%%%%%%%
\begin{abstract}
In this paper we present a fully relativistic study of axisymmetric magnetohydrodynamic Bondi--Hoyle accretion onto a moving Kerr black hole. The equations of general relativistic magnetohydrodynamics are solved using high resolution shock capturing methods. In this treatment we consider the ideal MHD limit. The parameters of interest in this study are the adiabatic constant $\Gamma$, the asymptotic speed of sound $c_{s}^{\infty}$, and the plasma beta parameter $\beta_{P}$. We focus the investigation on the parameter regime in which the flow is supersonic, or when $v_\infty \ge c_{s}^{\infty}$. In some cases, subsonic asymptotic flows are considered for comparison purposes. We study the accretion rates of the total energy and momenta, as well as the hydrodynamic energy and momentum accretion rates. The models presented in this study exhibit a matter density depletion in the downstream region of the black hole which tends to vacuum $(\rho_0=0)$ in convergence tests. This feature is due to the presence of the magnetic field, more specifically the magnetic pressure, and is not seen in previous purely hydrodynamic studies.
\end{abstract}
%%%%%%%%%%%%%%%%%%%%%%%%%%%%%%%%%%%%%%%%%%%%%%%%%%%
%\pacs{???} % NEEDS TO BE CHECKED!!!
%%%%%%%%%%%%%%%%%%%%%%%%%%%%%%%%%%%%%%%%%%%%%%%%%%%
\maketitle
%%%%%%%%%%%%%%%%%%%%%%%%%%%%%%%%%%%%%%%%%%%%%%%%%%%
%
%%%%%%%%%%%%%%%%%%%%%%%%%%%%%%%%%%%%%%%%%%%%%%%%%%%
%
%%%%%%%%%%%%%%%%%%%%%%%%%%%%%%%%%%%%%%%%%%%%%%%%%%%
\section{Introduction}
%%%%%%%%%%%%%%%%%%%%%%%%%%%%%%%%%%%%%%%%%%%%%%%%%%%

The Bondi--Hoyle--Lyttleton accretion process was originally studied in 1942 \cite{BondiHoyle}. This model of the accretion process assumes a massive point particle travels through a perfect fluid background (no heat transport) with a Newtonian treatment of gravity. Such a configuration is thought to approximate the dynamics of two bodies in a common envelope, or the dynamics of a body travelling through the medium accreting onto an active galactic nuclei \cite{BHL_Review}. In such astrophysical systems we are able to safely neglect viscosity in the fluid treatment due to the length and velocity scales involved in the dynamics. However, past investigations of this phenomenon rarely account for the presence of magnetic fields. In many cases this is due to the relative simplicity of the hydrodynamic models. Astrophysically, it is expected that material accreting onto, say, an AGN, will be highly ionized, and consequently a magnetic field will be generated by the accreting material \cite{bob}. Due to the high conductivity of the accretion disk, the magnetic field is bound to the disk, by an effect known as flux freezing \cite{Kulsrud}. The flux freezing is a characteristic of ideal magnetohydrodynamics. The typical magnetic field strengths for AGN have a wide range range, from $10^4$G \cite{Field:1993} to $10^{10}$G \cite{Kardashev:1995} (as cited by Tyul'bashev \cite{Tyulbashev:2002}). The magnetic fields observed in common envelopes are on the order of $10^7-10^8$G \cite{Sing:Thesis}.

% The Newtonian Bondi--Hoyle accretion phenomenon has been investigated by several authors over the years; recent investigations include \cite{FR1,FR2} and more recently \cite{FGR1}. Some of the more recent configurations under investigation include accretion with a constant vortex background \cite{Krumholz1, Krumholz2}. 

% The hydrodynamic Bondi problem has been used as a test case for general relativistic high resolution shock capturing hydrodynamic codes, in several papers by for example Hawley \etal~\cite{Hawley, VH1}. GR

We investigate a modified version of this problem where the massive point particle is replaced with a black hole with a non-trivial radius. Consequently, we must consider a relativistic treatment of gravity. This line of research began with Petrich \etal~\cite{Petrich:1988} where they found a closed form solution of an ultrarelativistic spherically symmetric black hole and found that the flows will be steady. Further work by the same authors, \cite{PSST}, solved the fluid equations of motion using numerical methods. In \cite{PSST} they studied both the point mass model with Newtonian gravity, as well as a model in which the point mass is replaced with a spherically symmetric black hole and relativistic treatment of gravity. Both the Newtonian and relativistic models were studied using axisymmetry. In both works they found that the flows were steady to long term evolution. Font \etal~\cite{FI1, FI2, PF1} extended this research to include an axisymmetric black hole, and in \cite{FI3} they modelled a infinitely thin disk model, where accretion is assumed to occur only in the equatorial plane. In all previous research only a hydrodynamic background fluid was studied, and all relativistic flows were determined to reach a steady state. This problem continues to be of interest in modern research such as in the work by Farris \etal~\cite{Farris:2009} who found an application of the three dimensional relativistic Bondi--Hoyle accretion in binary neutron star mergers, in perturbations of fully three dimensional hydrodynamic models for Bondi--Hoyle--Lyttleton accretion \cite{Blakely:Thesis}, and most recently in the search for QPO's once again using the infinitely thin disk approximation \cite{Donmez:2010}.

Our contribution to this research determines the phenomenology of the same system when including a background magnetic field. We extend the original analysis by Font \etal~\cite{FI1, FI2, PF1} by introducing an asymptotically uniform magnetic field as described by Wald \cite{Wald:1974}. In this paper, we use the ideal magnetohydrodynamic stress-energy tensor which reflects the presence of an embedded magnetic field. To parameterize the magnetic field, we use a method from plasma physics, which introduces the plasma beta parameter, $\beta_P$ \cite{Hawley:1995, Hawley:1996, Minuzo:2004}. The plasma beta parameter is a ratio of the magnetic pressure to the thermodynamic pressure. The ideal MHD approximation is only valid for systems in which the plasma beta parameter obeys the condition $\beta_{P}\lesssim1$. We use initial parameters for our system such that the plasma beta parameter is less than or approximately one over the entire domain of integration. As reported in \cite{FI1}, the interesting dynamics are in what is referred to as the supersonic regime, where the asymptotic velocity $v_\infty$ of the central body is greater than the asymptotic speed of sound $c_{s}^{\infty}$ in the fluid.

During this study we restrict our attention to the ``hot'' relativistic equation of state, $P=(\Gamma-1)\rho_0\epsilon$. In particular, we focus on two different values of the adiabatic constant $\Gamma$, one in the non-relativistic regime $\Gamma=5/3$, and the other the relativistic equation of state $\Gamma=4/3$. Although other values in between these are possible and considered physically valid, we feel that the interesting dynamics for a first approach to solving the GRMHD Bondi--Hoyle problem are captured by studying the extremes. In a future study we will study a wider range of the parameters.

One of the difficulties in studying magnetized fluids involves the enforcement of the $\nabla\cdot\bd{B}=0$ constraint. The numerical treatment of magnetic fields has only become tractable with the techniques developed such as the constraint transport method \cite{HawleyEvans:1988}, the flux transport method \cite{Toth:2000}, or the hyperbolic-divergence cleaning method \cite{Dedner:2002}. In this study we implement the hyperbolic divergence cleaning method as described by Palenzuela \etal~\cite{Palenzuela:2008}.

The outline of the paper is as follows: section \ref{Sec:0} will discuss the coordinates used in the problem. Section \ref{Sec:1} will discuss the equations of motion used for the relativistic ideal magnetohydrodynamic system. Section \ref{Sec:2} will discuss the initial conditions and boundary conditions used to perform the simulations. Section \ref{Sec:3} will briefly cover the numerical methods developed and used for the simulations. Section \ref{Sec:4} will discuss the flow morphology and in section \ref{Sec:5} we draw our conclusions.

In the rest of this paper we use geometric units, where $G=c=1$ with $c$ the speed of light in vacuum, and $G$ being Newton's gravitational constant. Further we will use the notation that Greek scripts run over the entire spacetime, i.e.~$\mu=0,1,2,3$, and Latin scripts run over the spatial coordinates, i.e.~$i=1,2,3$. All variables are assumed to be functions of both the radial and polar coordinates.

%+++++++++++++++++++++++++++++++++++++++++++++++++++++++++++++++++
\section{Coordinates\label{Sec:0}}
%+++++++++++++++++++++++++++++++++++++++++++++++++++++++++++++++++
In our study we are interested in the flow around a rotating black hole. Thus the line element defined by our black hole will be described using the Kerr spacetime. The original formulation of this line element contains coordinate singularities as is described in for example \cite{Camenzind}. To circumvent the coordinate singularity, we use what are know as Kerr--Schild coordinates, as described by Font \etal~\cite{PF1}. 
%Using these coordinates, particles may cross the event horizon without unphysical numerical density and pressure build-up. 
In the Kerr--Schild coordinate system, the rotating black hole line element is written
\begin{align}
ds^2=&-\left(1-\frac{2Mr}{\Delta}\right)dt^2+\frac{4Mr}{\Delta}dtdr\nonumber\\
&+\left(1+\frac{2Mr}{\Delta}\right)dr^2-\frac{4aMr\sin{\theta}^2}{\Delta}drd\phi\nonumber\\
&+(r^2+a^2\cos{\theta}^2)d\theta^2\nonumber\\
&+(\Delta+a^2\left(1+\frac{2Mr}{\Delta}\right)\sin{\theta}^2)\sin{\theta}^2d\phi^2,
\end{align}
with
\begin{equation}
 \Delta = r^2+a^2\cos{\theta}^2.
\end{equation}
$a$ is the dimensionless measure of the rotation rate of the black hole and is related to the angular momentum of the black hole via $J=M^2a$, where $M$ is the mass of the black hole. For the present study we set $M=1$ without loss of generality. %This is the simplest choice of the family of coordinates that describe the Kerr spacetime without coordinate singularities. For more discussion of the general family we refer the reader to \cite{PF1} or \cite{Camenzind}.
%When the black hole does not have angular momentum, $a=0$, the metric reduces to the ingoing Eddington--Finkelstein coordinates, taking the form
%
% \begin{align}
% ds^2=&-\left(1-\frac{2M}{r}\right)dt^2+\frac{4M}{r}dtdr\\
% &+\left(1+\frac{2M}{r}\right)dr^2+r^2d\Omega^2,
% \end{align}
%
%where $d\Omega^2=d\theta^2+\sin(\theta)^2d\phi^2$. 
% In this coordinate system, time and azimuthal angle have different meanings than those in the Schwarzschild coordinates. The coordinate transformation has the simple expression \cite{Krolik:2009};
% \begin{eqnarray}
% dt_{KS} &=& dt_{BL} + \frac{2Mr}{\Delta}dr\\
% dr_{KS} &=& dr_{BL}\label{eq:radial}\\
% d\phi_{KS} &=& d\phi_{BL} + \frac{a}{\Delta}dr\\
% d\theta_{KS} &=& d\theta_{BL}
% \end{eqnarray}
% 
% The conversion between the two coordinates is straightforward, but can only be done outside of the event horizon, which in both coordinates are located at the same radial coordinate \cite{PF1}.

%The remaining singularity, at $r=0$, is a consequence of the geometry and therefore is a physical singularity. As long as the rotating black hole satisfies $|a|\le1$ we will have an event horizon located at $(1-2Mr/\Delta)=0$ and thus the $r=0$ singularity is hidden. Our interest is only in potentially observable phenomenon and as such we restrict our attention to the region along and outside the event horizon.

In this work, we consider an axisymmetric spacetime geometry. As a result of this symmetry we restrict our study to an asymptotically uniform magnetic field which is aligned with the axis of rotation of the axisymmetric black hole. This particular magnetic field configuration is the only configuration that is compatible with the symmetries imposed. For more general magnetic field configurations, such as those presented in Bi\v{c}\'{a}k \etal~\cite{Bicak:1985}, we will require the use of a three dimensional code, currently under investigation.
%
%
%+++++++++++++++++++++++++++++++++++++++++++++++++++++++++++++++++
\section{Equations of Motion\label{Sec:1}}
%+++\++++++++++++++++++++++++++++++++++++++++++++++++++++++++++++++
To obtain the equations of motion for the ideal relativistic magnetohydrodynamic (MHD) system, we consider the conservation of the baryon density, the stress energy tensor and the Maxwell equations as seen in \cite{Camenzind}. To close the system of equations, we use an equation of state to relate the internal energy density to the thermodynamic pressure. To decompose our system of equations into the $3+1$ formalism we use the ADM $3+1$ variables with the spacetime metric
% \begin{eqnarray}
% \nabla_\mu J^{\mu} &=&0\label{eq:current}\\
% \nabla_\mu T^{\mu\nu}&=&0\label{eq:matter}\\
% \nabla_\mu \preup{*}{F}^{\mu\nu}&=&0\label{eq:Maxwell}\\
% P&=&(\Gamma-1)\rho_o\epsilon
% \end{eqnarray}
% where $J^{\mu}=\rho_o u^{\mu}$ is the current density of the fluid, $\rho_o$ is the specific rest mass density of the fluid particles, and $u^{\mu}$ is the fluid's 4-velocity. The quantity $T^{\mu\nu}$ is the magnetohydrodynamic stress-energy tensor which will be described in greater detail in section \ref{sec:SET}. $\preup{*}{F}^{\mu\nu}$ is the dual tensor to the Faraday tensor of electromagnetism \cite{Jackson}. $P$ is the fluid pressure, $\epsilon$ is the internal energy density of the fluid and $\nabla_\mu$ is the covariant derivative.  $\Gamma$ is the adiabatic constant of the relativistic polytropic equation of state of the fluid.
%
%\subsection{ADM $3+1$ Variables}\label{sec:ADM}
%We use the $3+1$ Arnowitt--Deser--Misner (ADM) \cite{York} variables to reformulate the spacetime metric into pieces that allow us to separate space and time. The ADM variables allow us to re-express the system of relativistic equations as a Cauchy problem which may be solved using standard hyperbolic partial differential equation techniques. This formulation records the spacetime metric as
\begin{equation}
 ds^2 = -(\alpha^2-\beta^i\beta_i)dt^2+2\beta_idtdx^i+\gamma_{ij}dx^idx^j
\end{equation}
where $\alpha$ is the lapse function, $\beta^i$ is the shift vector, and $\gamma_{ij}$ is the induced metric on the spacelike hypersurfaces.
%The transformation from spacetime variables to $3+1$ variables requires that we apply a projection operator to the spacetime variables, producing a set of variables that are parallel to the spacelike surfaces, or perpendicular to them. We denote this operator $\gamma_{\mu\nu}$,
%\begin{equation}
%\gamma_{\mu\nu} = g_{\mu\nu} + n_\mu n_\nu
%\end{equation}
%where $n^\mu$ is a 4-vector orthogonal to the spacelike hypersurfaces. For more detail about the $3+1$ decomposition we refer the reader to standard texts such as Misner \etal~\cite{MTW}.

\subsection{Hyperbolic Divergence Cleaning}
The evolution of the magnetic field has a physical constraint $\nabla\cdot B=0$, we have three evolution equations and one constraint equation. This leaves us with an over determined set of equations. Traditionally, the numerical treatment of a constrained system uses free evolution, where the evolution equations are used to evolve the system of equations and there is an implicit assumption that the constraint will be maintained. However, when using free evolution, any numerical errors that arise are often linked to constraint violations, thus we need a method that enforces the constraint as the flow evolves.

To maintain the magnetic field constraint we use the hyperbolic divergence cleaning method as originally proposed by Dedner \cite{Dedner:2002}, and used by other groups such as Neilsen \etal~\cite{Neilsen:2006,Anderson:2006}, and Palenzuela \etal~\cite{Palenzuela:2008}.

To implement the diffusive hyperbolic method, we add a divergence term acting on an auxiliary field $\psi$ of the form $\nabla_\mu(g^{\mu\nu}\psi)$ to the Maxwell equations $\nabla_\mu \preup{*}{F}^{\mu\nu}=0$,
\begin{equation}
\nabla_\mu (\preup{*}{F}^{\mu\nu}+g^{\mu\nu}\psi)=0\label{eq:modF}
\end{equation}
We also add a diffusive term, $\kappa n^{\mu}\psi$ to \eqref{eq:modF} to damp out any $\nabla\cdot B = 0$ violations \cite{Palenzuela:2008}, where
$\kappa$ is a tunable parameter. Our final expression for the Maxwell equations becomes
\begin{equation}
\nabla_\mu (\preup{*}{F}^{\mu\nu}+g^{\mu\nu}\psi)=\kappa n^{\mu}\psi\label{eq:modF2}
\end{equation}
In the absence of any $\nabla\cdot B = 0$ violations we expect that the extra parameter $\psi$ will reduce to zero and thus we recover the original formulation of the ideal MHD equations as described in \cite{Camenzind} for example.

\subsection{Stress-Energy Tensor}\label{sec:SET}
We use the now standard form of the ideal magnetohydrodynamic stress-energy tensor as described in, for example, Noble \etal~\cite{Noble}
\begin{equation}
T^{\mu\nu} = (\rho_o h+b^2)u^\mu u^\nu +(P+b^2/2)g^{\mu\nu}-b^{\mu}b^{\nu}
\end{equation}
where $h$ is the enthalpy of the system
\begin{eqnarray}
h = 1 + \epsilon + \frac{P}{\rho_0},
\end{eqnarray}
and $b^2$ is
\begin{equation}
 b^2 = g_{\mu\nu}b^{\mu}b^{\nu},
\end{equation}
where
\begin{equation}
 b^{\mu}=(W\gamma_{ij}B^{i}v^{j},B^i/W+\hat{v}^i).
\end{equation}
\subsection{Plasma Beta Parameter}
We introduce an expression to allow us to parameterize the magnetic field relative to the hydrodynamic contributions. The relativistic definition of the plasma beta parameter \cite{Hawley:1995,Hawley:1996} is 
\begin{equation}
 \beta_P = \frac{2P}{b^2}.
\end{equation}
When the magnetic field strength increases this parameter decreases, and when the magnetic field strength decreases this parameter increases. The magnetic fields in our simulations will be initialized using the asymptotic plasma beta parameter, $\beta_P^{\infty}$, which will fully specify the magnetic field strength.
\subsection{Equations of Motion}
We use the Valencia formulation for ideal MHD. These are most readily found in \cite{LivingRevFont,LivingRevMarti} thus we will not restate them here. We introduce 4 new variables relating to the divergence cleaning we implement, which are explained below.
\subsection*{New Variables}
Considering our hyperbolic divergence cleaning method we have two additional conserved quantities, the conserved divergence cleaned magnetic density, $\Pi^k$, and the divergence violation field, $\Psi$,
 \begin{align}
  \Pi^j &= \mathcal{B}^j+\beta^j\Psi\\
  \Psi  &= \psi/\alpha.\label{eq:new_con}
 \end{align}
We also have the corresponding primitive variables,
\begin{eqnarray}
 \mathcal{B}^j &=& \Pi^j - \beta^j\Psi\\
 \psi &=& \alpha \Psi.\label{eq:new_prim}
\end{eqnarray}
Just as in the standard GRMHD model, there are no known closed form solutions to the inverse relations, so we use numerical methods to perform this conversion when necessary. Primitive variable recovery is performed using a modified version of Mignone's one parameter inversion scheme \cite{Mignone:2006}. This was chosen due to its simplicity, and when we compared this inversion scheme against the two variable solver promoted by Noble \etal~\cite{Noble} there was no performance difference. The scheme is readily found in \cite{Mignone:2006} and is not repeated here. The primitive recovery for variables $\mathcal{B}^j$ and $\psi$ are trivial and are seen in Eqn.~\eqref{eq:new_prim}.

\subsection*{Modified Equations of Motion}
To determine the equations of motion for the new variables we take the $3+1$ projection of Eqn.~\eqref{eq:modF2}, to get
 \begin{eqnarray}
  \partial_t \sqrt{\gamma} \Pi^j &+& \frac{\partial}{\partial x^i} \sqrt{-g}\left(\mathcal{B}^j\hat{v}^i-\mathcal{B}^i\hat{v}^j+\alpha g^{ij}\Psi\right)\nonumber\\ 
   &=& -\alpha\sqrt{-g}g^{\alpha\mu}\Gamma^j_{\alpha\mu}\Psi+\kappa\beta^j\Psi\label{eq:con4}\\
 \partial_t \sqrt{\gamma} \Psi &+& \frac{\partial}{\partial x^i} \sqrt{-g}\left(\frac{\mathcal{B}^i}{\alpha}-\frac{\beta^i}{\alpha}\right)\nonumber\\
  			     &=& -\alpha\sqrt{-g}g^{\alpha\mu}\Gamma^t_{\alpha\mu}\Psi+\kappa\Psi.\label{eq:con5}
 \end{eqnarray}
The remaining equations of motion are as presented in \cite{LivingRevFont}.
\section{Initialization and Boundary Conditions\label{Sec:2}}
%+++++++++++++++++++++++++++++++++++++++++++++++++++++++++++++++++
We use the method described by Font \etal~\cite{FI1,FI2} to initialize the hydrodynamic variables. Since the characteristics of the system are independent of the choice of initial pressure, we use $P_\infty = 1$ and note that this choice is entirely arbitrary. The velocity fields are initialized using the following form \cite{FI1}:
\begin{eqnarray}
v^r &=& \frac{1}{\sqrt{g_{rr}}}v_\infty\cos{\theta}\nonumber\\
v^\theta &=& -\frac{1}{\sqrt{g_{\theta\theta}}}v_\infty\sin{\theta}\\
v^\phi &=& 0\nonumber.
\end{eqnarray}
One may easily verify that $v^2=v_\infty^2$.

The magnetic field is initialized using Wald's solution \cite{Wald:1974},
\begin{equation}
 F_{\mu\nu} = B_0\left(\preup{(\phi)}{\xi}_{[\mu,\nu]}+2a\preup{(t)}{\xi}_{[\mu,\nu]}\right).
\end{equation}
$\preup{(\phi)}{\xi}_{\mu}$ and $\preup{(t)}{\xi}_{\mu}$ are the azimuthal and temporal Killing vectors respectively, the square brackets denote anti-symmetrization, and $a$ is the spin parameter of the black hole. $B_0$ is a scaling factor that determines the magnitude of the magnetic field. This reduces to the initial conditions for the magnetic field components;
\begin{eqnarray}
 B^r      = \frac{-B_0}{2\sqrt{\gamma}}\left(\gamma_{\phi\phi,\theta}+2ag_{\phi t,\theta}\right)\\
 B^\theta = \frac{-B_0}{2\sqrt{\gamma}}\left(\gamma_{\phi\phi,r}+2ag_{\phi t,r}\right)\\
 B^\phi   = \frac{-B_0}{2\sqrt{\gamma}}\left(\gamma_{r\phi,\theta}+2ag_{r t,\theta}\right)
\end{eqnarray}
This configuration is divergence free, and is uniform as $r\rightarrow\infty$. Due to the symmetry of our configuration we align the magnetic field with the rotation axis of the black hole.

The domain of integration for this study is defined by $r_{\min} \le r \le r_{\max}$ and $0\le\theta\le\pi$. $r_{\min}$ is determined in such a way that it will always fall inside the event horizon. $r_{\max}$ was set to be sufficiently far from the event horizon that it would be effectively considered infinity. For some simulations, if the radial domain is not set to be large enough, unphysical waves travel back towards the black hole, ultimately destabilizing the system.

The boundary conditions for this problem are broken into three regions, two for the radial coordinate, and one for the angular.
\begin{enumerate}[label=(\roman{*})]
 \item In the radial direction, near the event horizon, the flow is strictly absorbing, also known as outflow. In our implementation all scalar quantities are copied to the ghost cell regions and all vector quantities in the radial direction are linearly extrapolated to the ghost cells. This mixing of methods prevented unphysical backward traveling waves exiting the event horizon into the rest of the domain from destroying the setup.
 \item For the boundary at the outer radial domain, we use prescribed boundaries ``upstream'' of the black hole, and outflow boundaries ``downstream'' of the black hole. Upstream was determined to be any region with angular coordinate in the range $\pi/2\le\theta\le\pi$, while the remaining domain is considered downstream. The inflow condition is a prescribed boundary condition where we use the same asymptotic values as defined in the initialization routine.
 \item In the angular direction, due to the symmetries considered in this study, reflective boundary conditions are used, which copy all scalar variables, and all non-$\theta$-component variables to ghost cells. This procedure inverts the sign of the $\theta$-component of all vector quantities.
\end{enumerate}
%
%
%+++++++++++++++++++++++++++++++++++++++++++++++++++++++++++++++++
\section{Numerical Methods\label{Sec:3}}
%+++++++++++++++++++++++++++++++++++++++++++++++++++++++++++++++++
The equations of motion for our system take on the general form,
\begin{equation}\label{eq:conservative}
\frac{\partial}{\partial t}\sqrt{\gamma}\:{\bf{Q}}+\frac{\partial}{\partial x^i}\sqrt{-g}\:{\bf{F}}^i({\bf{Q}}) = \sqrt{-g}\:{\bf{S}}(\bd{Q}),
\end{equation}
where ${\bf{Q}}$ are the conservative variables, ${\bf{F}}^i$ denotes the flux, and ${\bf{S}}$ are the geometric source terms.
To solve this system, we developed our own high resolution shock capturing MHD code based on algorithms presented in the works \cite{HS1, FI1, PF1, Neilsen, Noble, Palenzuela:2008}. These are all based on an integral solution of Eqn.~(\ref{eq:conservative}) which allows for discontinuities in each fluid variable. In the presence of a discontinuity these methods solve a one dimensional Riemann problem.

We use the second order variable reconstruction MC limiter, with the HLL (Harten, Lax, and van Leer) approximate Riemann solver. The system was time evolved using second order Runge--Kutta integration. It is noted that, like all codes using approximate Riemann solvers, near extremum and discontinuities the order of accuracy reduces to first order. For the results presented we used regular grid spacing with a $400\times160$ numerical grid.

% There are problems using approximate Riemann solvers, although they offer a higher order of accuracy than using a regular Riemann solver, they only approximate the solution near discontinuities. For example the Roe method has known problems near extreme shocks, as presented in Nielsen \cite{Neilsen}. The HLL methods are diffusive and fail to capture turbulence, as is seen in the test cases run by (Stone) where the Kelvin--Helmholtz instability is not captured when using such a method. One needs to make a judgment call about the system as to which method is most suitable. In this system, the shocks are particularly smooth, and only in the limit of high speed of sound does the flow appear unsteady. Through several trials it was determined that the HLL solver was sufficient to calculate the trends of this system.

Parallelization was performed using the PAMR infrastructure developed by Frans Pretorius \cite{PAMR} as well as direct use of the Message Passing Interface (MPI). Simulations were performed using the woodhen cluster at Princeton, USA, and the WestGrid cluster in Canada.
%
%

%+++++++++++++++++++++++++++++++++++++++++++++++++++++++++++++++++
% THE TABLE OF VALUES
%+++++++++++++++++++++++++++++++++++++++++++++++++++++++++++++++++
\begin{table}
\begin{tabular}{|c|c|c|c|c|c|}
\hline
Model & $\Gamma$ & $a$ & $v_\infty$ & $c_s^{\infty}$ & $O(\beta^{\infty}_P)$\\
\hline
M1  & $4/3$ & 0   & 0.5 & 0.1 & 2\\
M2  & $5/3$ & 0   & 0.5 & 0.1 & 1\\
M3  & $4/3$ & 0.5 & 0.9 & 0.1 & 2\\
M4  & $4/3$ & 0   & 0.5 & 0.1 & 3\\
M5  & $4/3$ & 0   & 0.5 & 0.1 & 4\\
M6  & $5/3$ & 0   & 0.5 & 0.1 & 2\\
M7  & $5/3$ & 0   & 0.5 & 0.1 & 3\\
M8  & $5/3$ & 0   & 0.5 & 0.1 & 4\\
M9  & $4/3$ & 0.5 & 0.5 & 0.1 & 3\\
M10 & $4/3$ & 0.5 & 0.5 & 0.1 & 4\\
H1  & $4/3$ & 0   & 0.5 & 0.1 & $\infty$\\
H2  & $5/3$ & 0   & 0.5 & 0.1 & $\infty$\\
H3  & $4/3$ & 0.5 & 0.5 & 0.1 & $\infty$\\
\hline
\end{tabular}
\caption{Table of parameters used for the axisymmetric systems studied in this paper. We perform a small sampling of the available parameter space. Only supersonic flows, $v_\infty > c_{s}^{\infty}$, are investigated. The ``H'' models are purely hydrodynamic models.}
\label{table:1}
\end{table}
%
%
%+++++++++++++++++++++++++++++++++++++++++++++++++++++++++++++++++
\section{Accretion Profiles\label{Sec:5}}
%+++++++++++++++++++++++++++++++++++++++++++++++++++++++++++++++++
In this section we describe the new and major features of each flow studied. This is done by presenting cross sections of the pressure accretion profiles.

In Fig.~\ref{fig:axiPcross} we see the cross section of the final state for model M3. The upstream profile at the boundary smoothly agrees with the the profile, indicating that boundary effects will not be an issue in these simulations. The effects due to the presence of a magnetic field are minimal in the upstream region, and the magnetic pressure is uniform except in the region closest to the black hole. The downstream profile reveals a different picture. As the black hole accretes, the magnetic field builds up, the effects are largest near the black hole. However, the effects of the magnetic field are noticeable well outside of the hydrodynamic accretion radius, $r_{\rm{acc}}$,
\begin{equation}
r_{\rm{acc}} \equiv \frac{M}{v_{\infty}^2+(c_{s}^{\infty})^2}\label{eq:racc}.
\end{equation}
$c_s^{\infty}$ is the speed of sound in the fluid as measured by an asymptotic observer, M is the mass of the accretor, and $v_{\infty}$ is the velocity of the accretor as measured by an observer at asymptotic infinity.
This radius is the relativistic extension to the Newtonian Bondi radius as explained in detail by Petrich \etal~\cite{PSST}. Matter outside this radius is subsonic, whereas matter within this radius has a supersonic flow and will inevitably fall into the black hole. We refer the reader to \cite{BHL_Review} for details of the Newtonian calculations and more history. For the relativistic flows, this radius is fairly small, typically $0.5 \le r_{\rm{acc}} \le 1.5$. For the sake of consistency, we choose to measure the accretion rate at the event horizon, where we can focus on the material as it falls into the black hole. The radial location of the event horizon only depends on the mass and the angular momentum of the black hole, so this is constant for each choice of fluid parameters.

\begin{figure*}
\begin{minipage}{7in}
\includegraphics[width=3in]{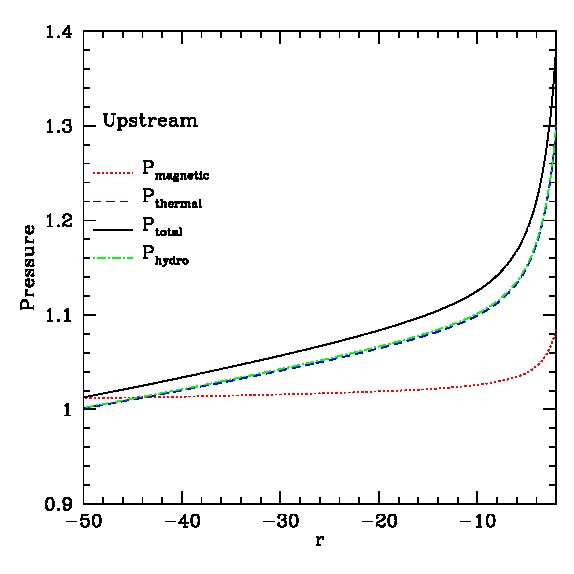}
\includegraphics[width=3in]{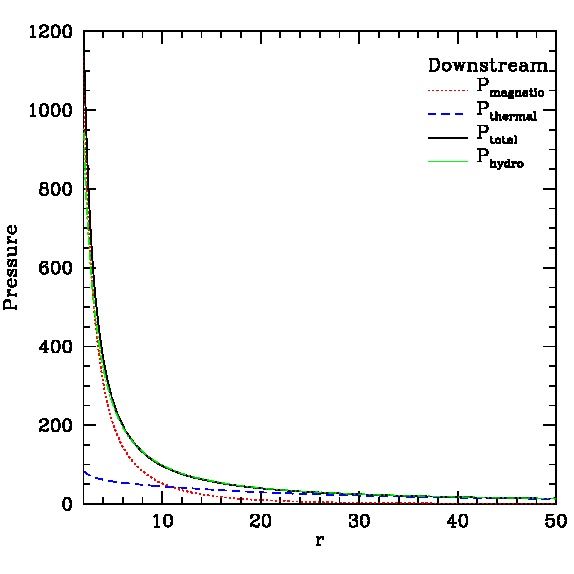}
\caption{We show the pressure cross-section in the upstream region (left) and downstream (right) along the axis of symmetry plotted from $r_{\rm{EH}}\le r \le 50$ for model M1. We see that the asymptotic behaviour of our system agrees with our boundary conditions, indicating that boundary effects are negligible for these simulations. We see a small effect due to the presence of the magnetic pressure in the upstream region; however, in the downstream region the magnetic effects are far more dominant in the region $r_{\rm{EH}}\le r \lesssim 10$ which is outside of the expected hydrodynamic accretion radius as defined in Eqn.~\eqref{eq:racc}. The upstream thermal pressure agrees with the hydrodynamic pressure from model H1, and the downstream total pressure profile agrees with the pressure profile of the same model. The magnetic pressure profile was shifted by one to fit the scale presented here. As is described in \cite{FI1} the upstream pressure cross section takes on the appearance of a Bondi accretor while the downstream region is far more extreme.}\label{fig:axi_G43_Pcross}
 \end{minipage}
\end{figure*}

\begin{figure*}
\begin{minipage}{7in}
\includegraphics[width=3in]{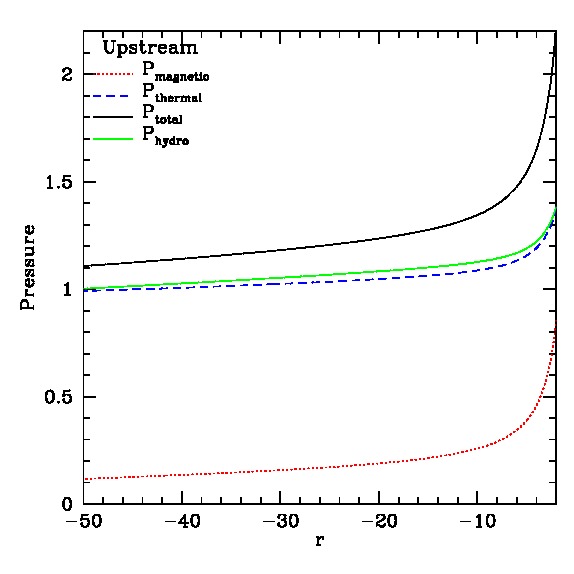}
\includegraphics[width=3in]{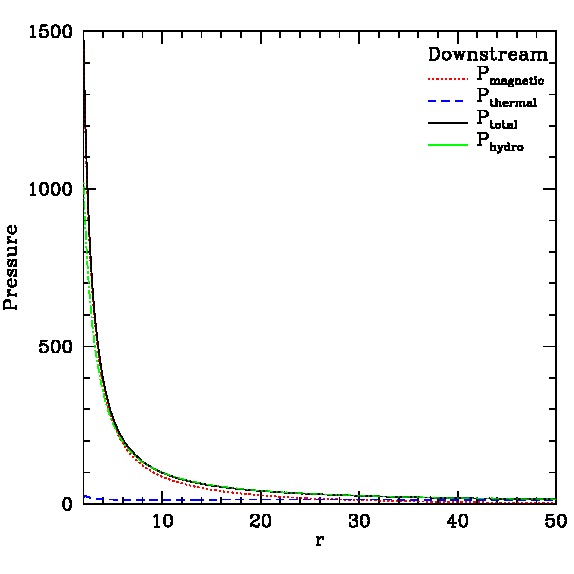}
\caption{We show the pressure cross-section in the upstream region (left) and downstream (right) along the axis of symmetry plotted from $r_{\rm{EH}}\le r \le 50$ for model M2. We see a similar profile as seen in Fig.~\ref{fig:axi_G43_Pcross}. In this model the magnetic pressure has a much larger amplitude upstream than for model M1, and shifting the data was not necessary. Upstream we also see a deviation between the hydrodynamic model and the thermal pressure. In the downstream region the magnetic effects are far more dominant in the region $r_{\rm{EH}}\le r \lesssim 20$ well outside the hydrodynamic accretion radius. The downstream the total pressure from model M2, and the hydrodynamic pressure from model H2 are again very similar.}\label{fig:axi_G53_Pcross}
 \end{minipage}
\end{figure*}

\begin{figure*}
\begin{minipage}{7in}
\includegraphics[width=3in]{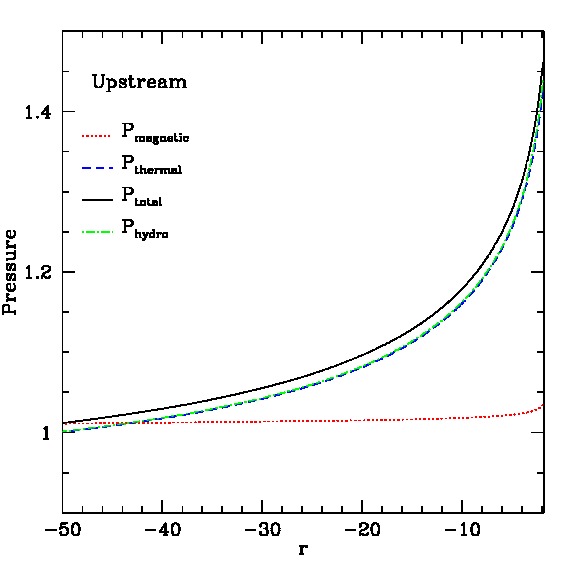}
\includegraphics[width=3in]{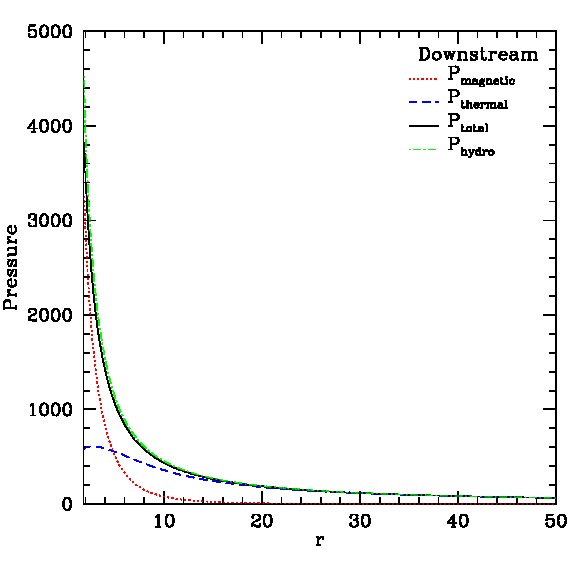}
\caption{We show the pressure cross-section in the upstream region (left) and downstream (right) along the axis of symmetry plotted from $r_{\rm{EH}}\le r \le 50$ for model M3. The profile in this model is very similar to that seen in Fig.~\ref{fig:axi_G43_Pcross}, including the upstream similarities between the hydrodynamic pressure and the thermal pressure. The effect of the rotating black hole is seen by the shortened magnetic pressure profile, which decreases to $r_{\rm{EH}}\le r \lesssim 6$ which, as in Fig.~\ref{fig:axi_G53_Pcross}, is outside of the expected hydrodynamic accretion radius. Careful inspection also shows that the downstream thermal pressure profile is dramatically different between models M1 and M3. The thermal pressure maximum occurs around $r=3M$ here whereas in model M1 the maximum is on the event horizon. Like in Fig.~\ref{fig:axi_G43_Pcross}, the magnetic pressure was shifted by one to fit the scale.}\label{fig:axiPcross}
 \end{minipage}
\end{figure*}

The profiles presented in Figs.~\ref{fig:axi_G43_Pcross}, \ref{fig:axi_G53_Pcross} and \ref{fig:axiPcross} have similar general appearance. The upstream pressure profile takes on the appearance of a stationary Bondi accretor, as was originally presented in Font \etal~\cite{FI1}. We find that the similarity continues when comparing the upstream profile of our models to the magnetic Bondi accretor. Just as in \cite{FI1} the downstream region is significantly different from the stationary accretor. In all cases the downstream profile indicates that the magnetic pressure builds up significantly in this region, and dominates the total pressure close to the event horizon of the black hole. An interesting trend occurs when modeling M1 and M3, the upstream profiles have the exact same trends; however, for model M3 in the downstream region the magnetic pressure is more than three times larger, and the effect has a shorter range. The profile for model M1 shows that the magnetic pressure extends to approximately $r=20M$, while the same profile for model M3 shows that the magnetic pressure only extends to approximately $r=15M$.

% \begin{figure}
% \includegraphics[width=3in]{pics/Gamma_43/v=0.5/vel_mag0002.pdf}
% \caption{Here we present the overlap between the magntitude of the fluid 3-velocity (blue-black lines) and the magnetic pressure (other colours) for model M1. We see that the region where the magnetic pressure dominates the velocity field  }\label{fig:axi7up}
% \end{figure}

% \begin{figure}
% \includegraphics[width=3in]{pics/Gamma_43/v=0.5/vel_mag0003.pdf}
% \caption{}\label{fig:axi7up2}
% \end{figure}

% \begin{figure}
% \includegraphics[width=3in]{pics/Gamma_43/v=0.5/rho_Pmag0000.pdf}
% \caption{We present the rest mass density (colour plot) and the magnetic pressure (contours) on a logarithmic scale for model M1. It is evident that the evacuation in the downstream side of the black hole is caused by a buildup of the magnetic pressure in the same region. These plots are produced using Visit. }\label{fig:axi7up3}
% \end{figure}

\begin{figure*}
\includegraphics[width=6in]{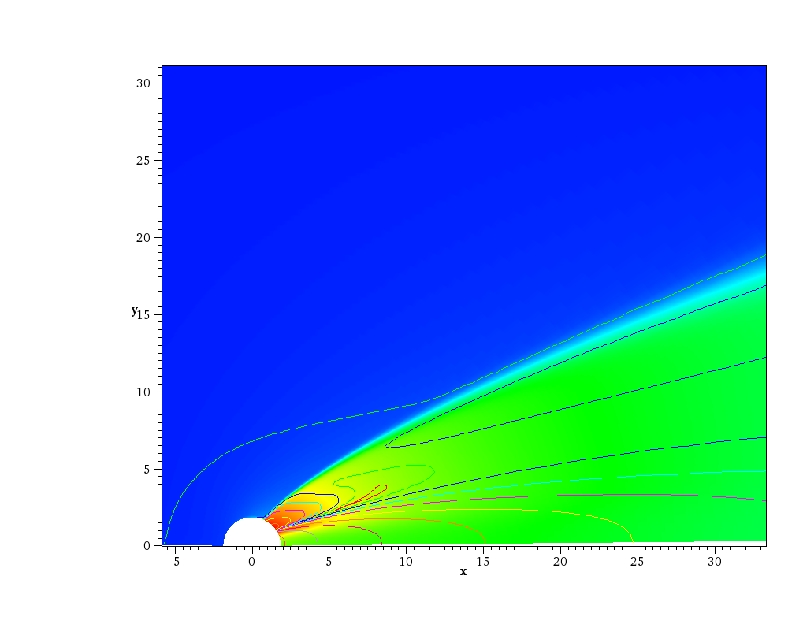}
\caption{The thermodynamic pressure (colour plot) and the magnetic pressure (contours) on a logarithmic scale for model M1. We see that there is a balance between the two sources of pressure. This is made more apparent in figure \ref{fig:all_P_M1}. We also see that the magnetic field wraps around the black hole as shown by the dark blue contour connecting the magnetic pressure downstream to the magnetic pressure upstream. The magnetic reconnection downstream is sufficiently weak that the magnetic field lines pile up on the downstream side.}\label{fig:axi7up4}
\end{figure*}

In Fig.~\ref{fig:axi7up4} we present a pseudocolour plot for the thermal pressure for model M1 with a contour plot of the magnetic pressure. %The magnetic field wraps around the black hole, and collects in the downstream region. 
The magnetic field wraps around the black hole, as is seen by the magnetic pressure contour in the upstream region that traces the edge of the Mach cone in the downstream region. This forces the matter to decrease there. The magnetic reconnection
%
%The magnetic pressure builds downstream of the black hole forcing the matter density to decrease. The magnetic field wraps around the black hole, as is seen by the magnetic pressure contour in the upstream region that traces the edge of the Mach cone in the downstream region. The magnetic reconnection
%
in the downstream region is sufficiently weak that the magnetic field lines pile up in the downstream region. The magnetic pressure forces the downstream matter away from the axis of symmetry creating an evacuated region, which is similar to the effect as seen in the plasma depletion region associated with the earth's magnetosphere.

\begin{figure*}
\includegraphics[width=6in]{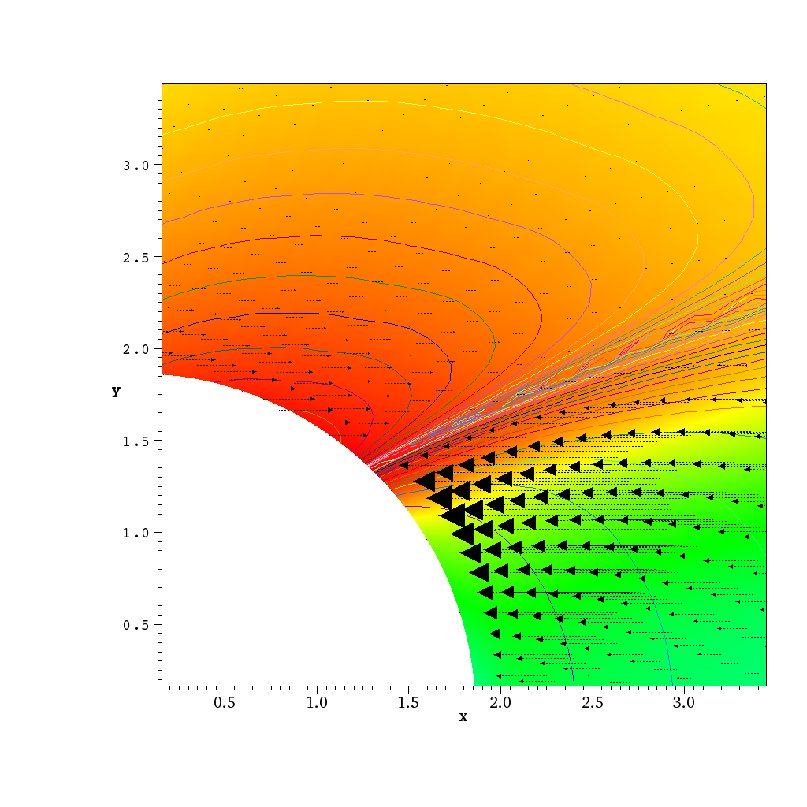}
\caption{The thermodynamic pressure (colour plot) and the magnetic pressure (contours) on a logarithmic scale for model M2. We superimpose the magnetic field vectors. We see that the downstream side has a distinct region where the magnetic field vectors diminish and switch direction. The thermal pressure balances the magnetic pressure in this region. This is a similar effect as seen in the plasma depletion region associated with the earth's magnetosphere.}\label{fig:mag_cont}
\end{figure*}

In Fig.~\ref{fig:mag_cont} we see the pseudocolour plot of the thermal pressure, with a contour plot of the magnetic pressure for model M2. In addition we present the magnetic field vectors. Outside of the Mach cone the magnetic field points in the same direction as the asymptotic magnetic field, however the field reduces magnitude, and reverses direction when it crosses into the tail shock. The magnitude of the magnetic field is represented by the size of the head of the vector. The magnitude is the largest closest to the black hole.

\begin{figure*}
\begin{minipage}{7in}
\includegraphics[width=2in]{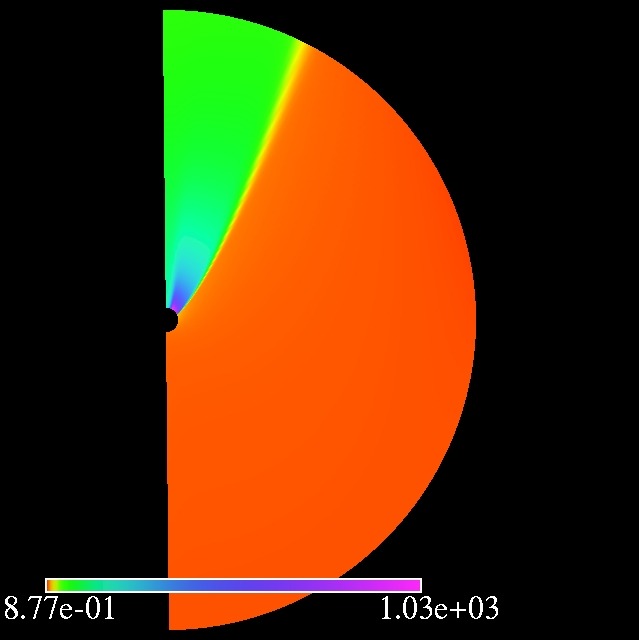}
\includegraphics[width=2in]{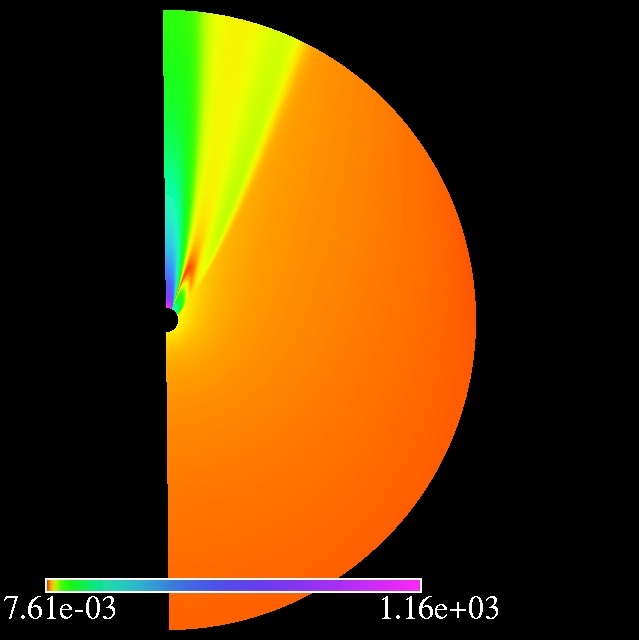}
\includegraphics[width=2in]{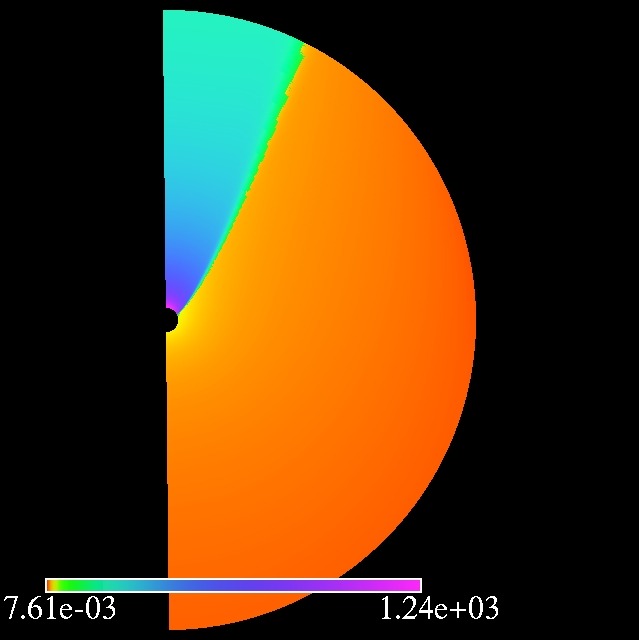}
\caption{The thermodynamic pressure (left), the magnetic pressure (middle) and the total pressure (right) for model M1. We see that the total pressure takes on a familiar form, exhibiting a tail shock as seen in the work by Font \etal~\cite{FI1}.}\label{fig:all_P_M1}
\end{minipage}
\end{figure*}

\begin{figure*}
\begin{minipage}{7in}
\includegraphics[width=2in]{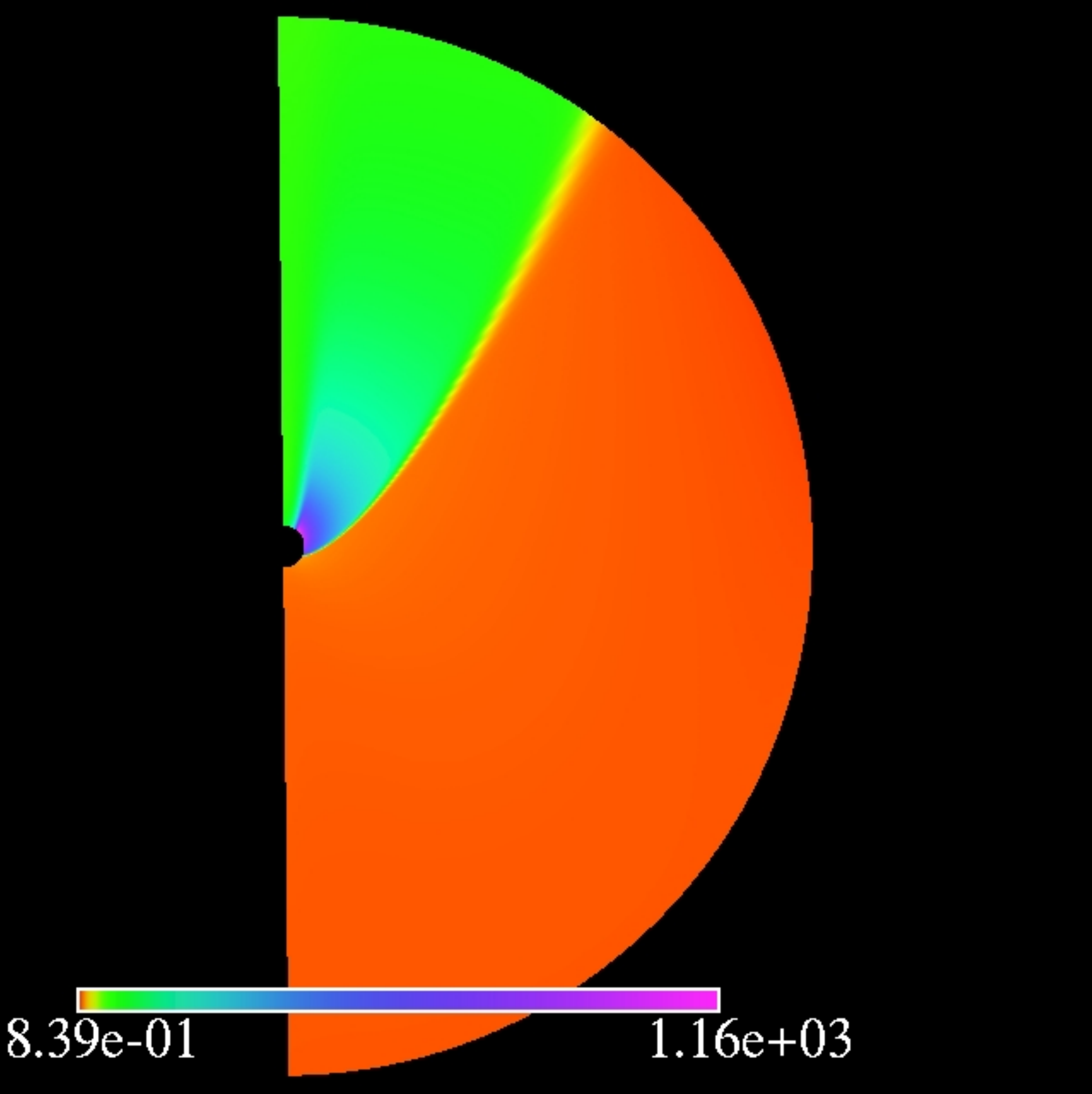}
\includegraphics[width=2in]{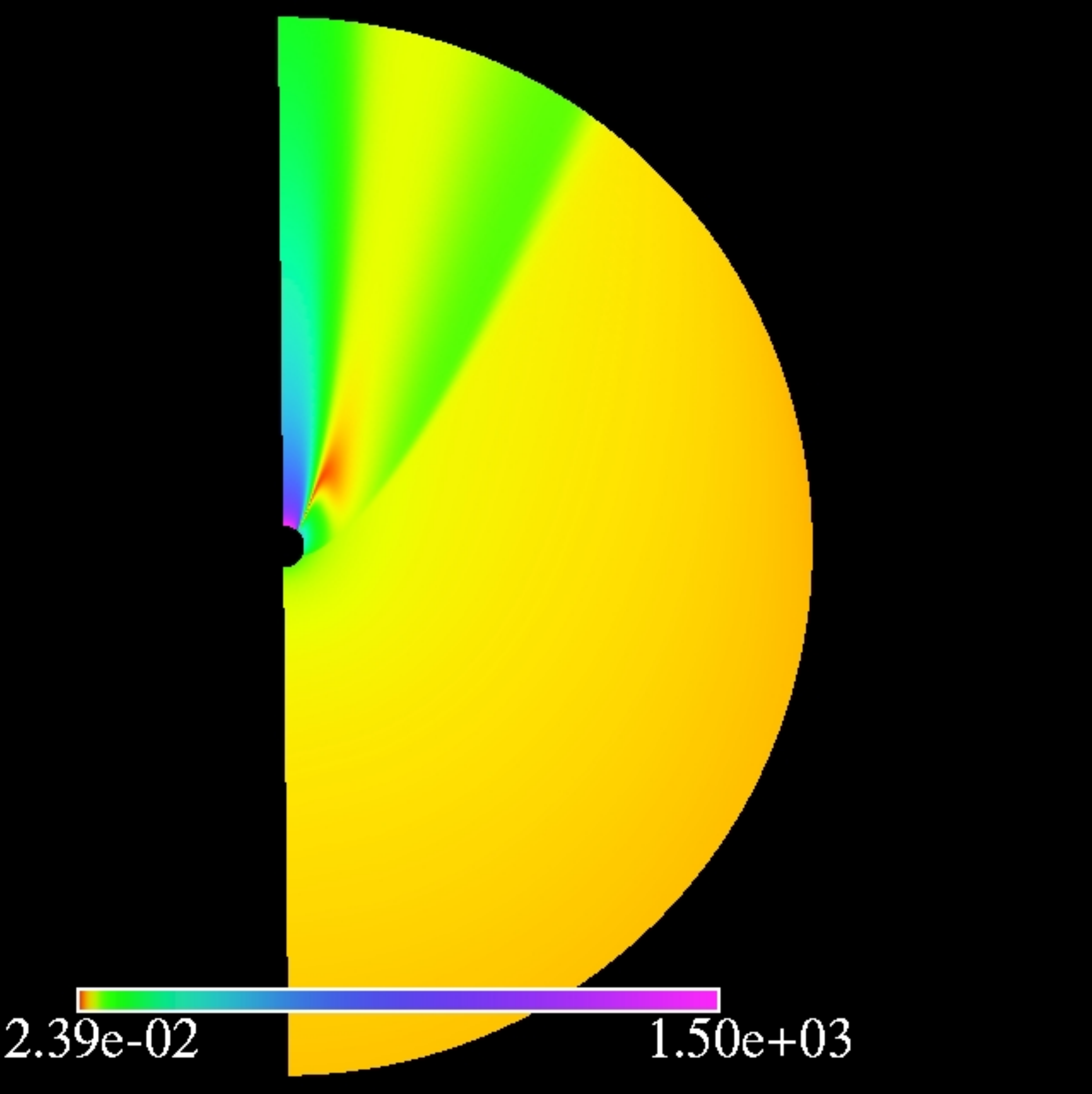}
\includegraphics[width=2in]{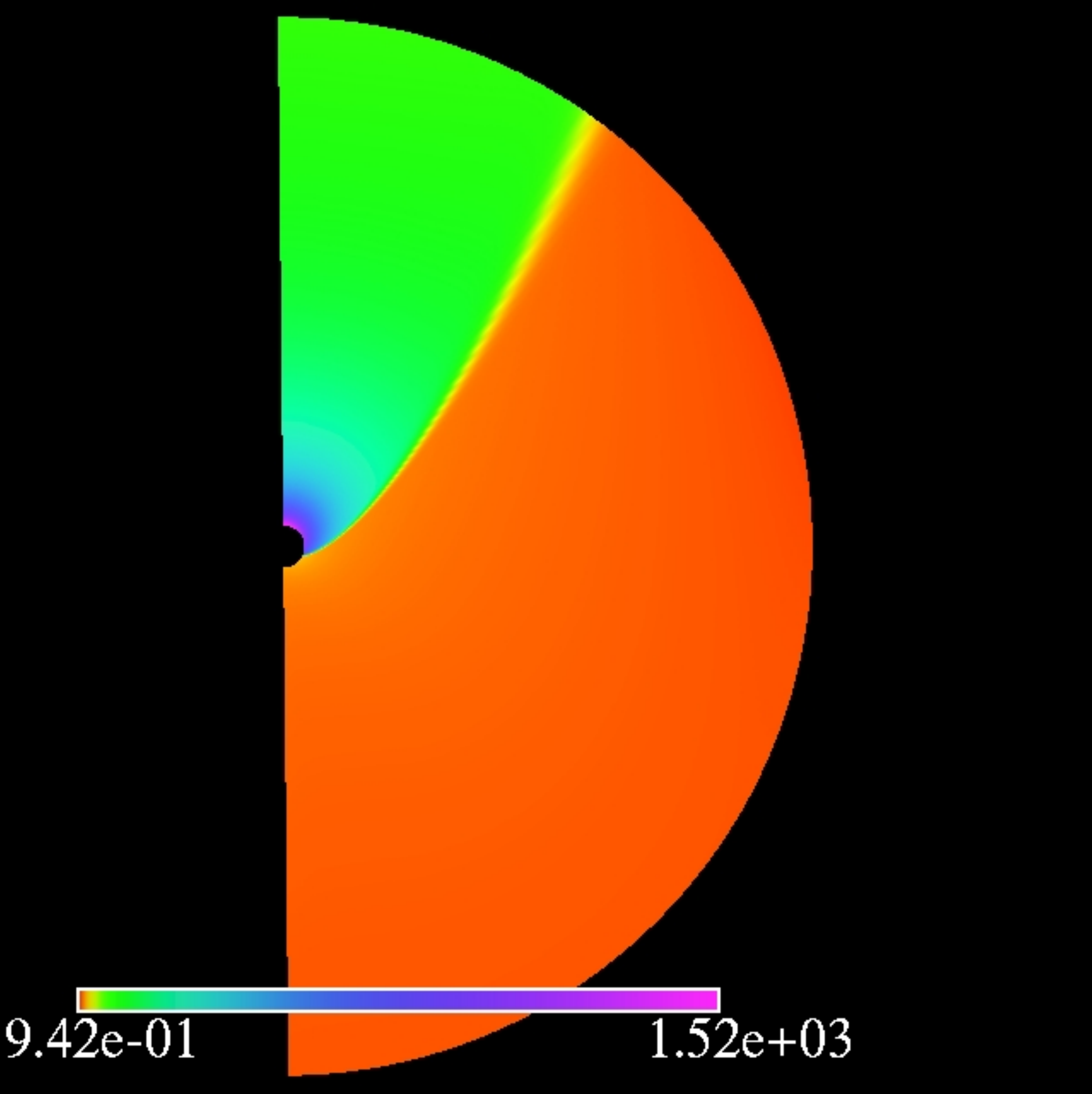}
\caption{The thermodynamic pressure (left), the magnetic pressure (middle) and the total pressure (right) for model M2. We see that the total pressure takes on a familiar form, exhibiting a tail shock as seen in the work by Font \etal~\cite{FI1}.}\label{fig:all_P_M2}
\end{minipage}
\end{figure*}

\begin{figure*}
\begin{minipage}{7in}
\includegraphics[width=2in]{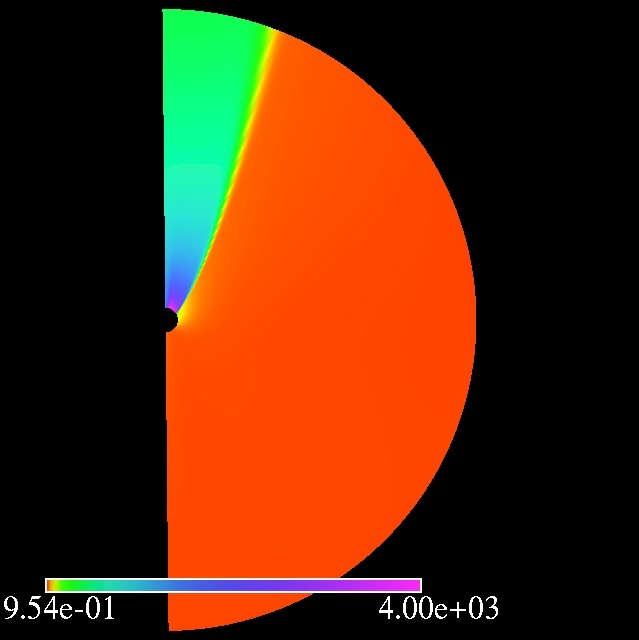}
\includegraphics[width=2in]{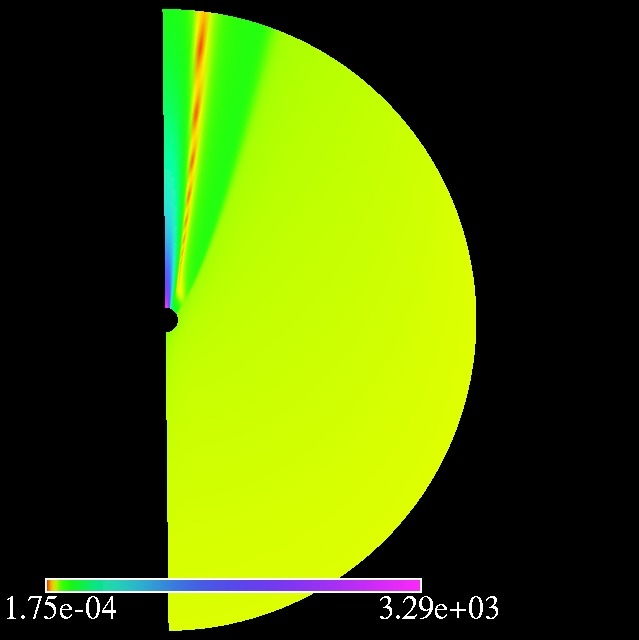}
\includegraphics[width=2in]{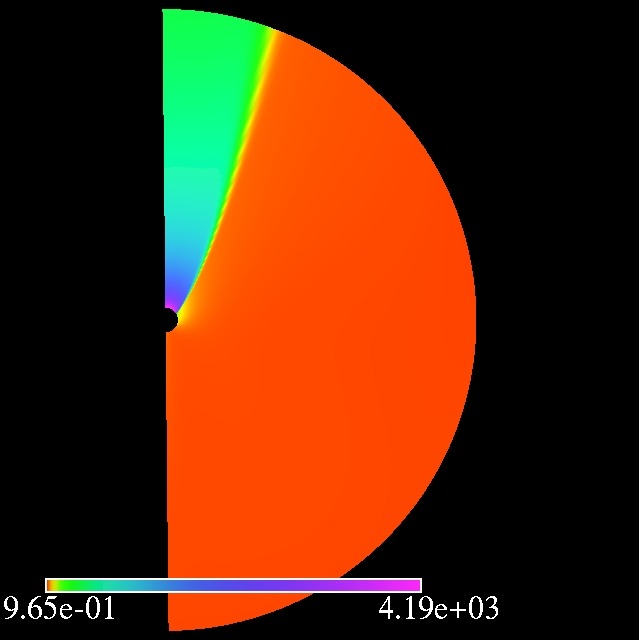}
\caption{We present the thermodynamic pressure (left), the magnetic pressure (middle) and the total pressure (right) for model M3. We see that the total pressure takes on a familiar form, exhibiting a tail shock as seen in the work by Font \etal~\cite{FI1}.}\label{fig:all_P_M3}
\end{minipage}
\end{figure*}

Figures \ref{fig:all_P_M1}, \ref{fig:all_P_M2}, and \ref{fig:all_P_M3}, present evidence that there is a balance between the thermal pressure and the magnetic pressure, in particular the total pressure has a similar profile to the hydrodynamic models.

A general feature of all relativistic Bondi--Hoyle accretion is the presence of a Mach cone \cite{FI1, FI2, FI3, Donmez:2010}. This cone attaches to the downstream side of the black hole with what is known as the opening angle. We observe the opening angle of the Mach cone, seen in Figs.~\ref{fig:all_P_M1},\ref{fig:all_P_M2}, and \ref{fig:all_P_M3}. The opening angle is known to be a function of the parameters of the fluid for Newtonian systems \cite{FR1}, and is seen in the study performed by Font \etal~\cite{FI1}.

\begin{figure}
\includegraphics[width=3in]{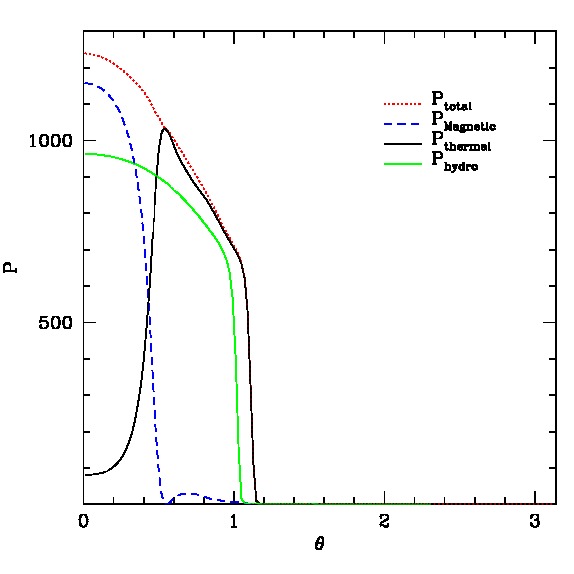}
\caption{A cross section at $r=2M$ for model M1 of the thermodynamic pressure, magnetic pressure, and the total pressure, as well as the pressure for model H1. The balance between the two sources of pressure is more clear in this cross section since the total pressure exhibits a similar form as those seen in purely hydrodynamic flows \cite{FI1}. The effect of the magnetic field increases the overall pressure in this cross section, especially along the axis of symmetry, and widens the Mach cone angle of attachment.}\label{fig:cross_P_M1}
\end{figure}

\begin{figure}
\includegraphics[width=3in]{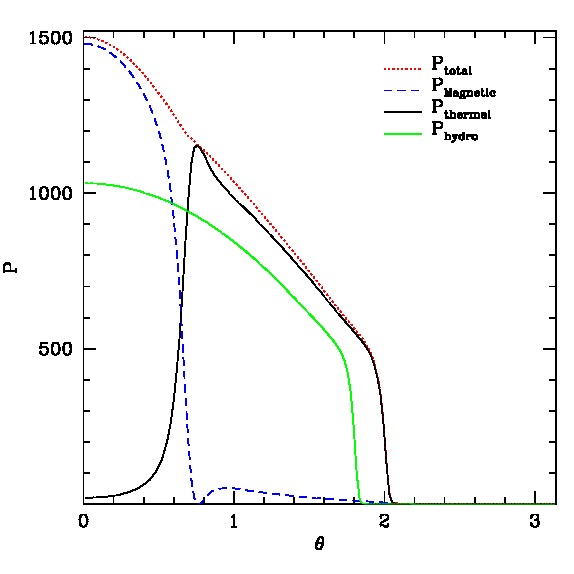}
\caption{A cross section at $r=2M$ for model M2 of the thermodynamic pressure, magnetic pressure, and the total pressure. We also plot the same cross section of the pressure for model H2. The balance between the two sources of pressure clearer in this cross section since the total pressure exhibits a similar form as those seen in purely hydrodynamic flows \cite{FI1}. As was the case in model M1, the presence of the magnetic field amplifies the downstream pressure along the axis of symmetry and widens the Mach cone angle of attachment. Since the region downstream, inside the Mach cone, is the location of the maximum accretion, we expect that the larger cross sectional area leads to an increased accretion rate, as is seen in Fig.~\ref{fig:ener_M1_M2}.}\label{fig:cross_P_M2}
\end{figure}

\begin{figure}
\includegraphics[width=3in]{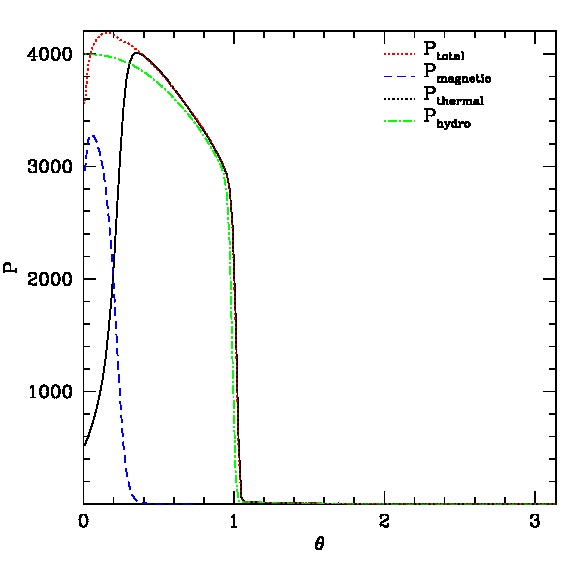}
\caption{A cross section at $r=2M$ for model M3 of the thermodynamic pressure, magnetic pressure, and the total pressure, and the pressure profile for model H3. The balance between the two sources of pressure is more clear in this cross section since the total pressure exhibits a similar form as those seen in purely hydrodynamic flows \cite{FI1}. As in the case for model M1, we find that the total pressure exceeds that for model H3, where no magnetic field is present; however, model H3 presents a maximum pressure on the axis of symmetry/rotation, whereas the total pressure for model M3 decreases on the axis. Further investigation is necessary to relate the spin of the black hole to the pressure decrease in the magnetic field cases.}\label{fig:cross_P_M3}
\end{figure}

In Figs.~\ref{fig:cross_P_M1}, \ref{fig:cross_P_M2}, and \ref{fig:cross_P_M3} we present the cross section of the pressure accretion profiles on $r=2M$ for models M1, M2, and M3, as well as models H1, H2, and H3 for comparison. The balance between the two sources of pressure, magnetic and thermal, is clear in these cross sections since the total pressure exhibits a similar form as those seen in purely hydrodynamic flows \cite{FI1}. The effect of the magnetic field increases the overall pressure in this cross section, especially along the axis of symmetry, and widens the Mach cone opening angle. The presence of the magnetic field amplifies the downstream pressure along the axis of symmetry and widens the Mach cone angle of attachment. Since the region downstream, inside the Mach cone, is the location of the maximum accretion, we expect that the larger cross sectional area leads to an increased accretion rate, as is seen in Fig.~\ref{fig:ener_M1_M2}. In Fig.~\ref{fig:cross_P_M3}, we see that model H3 presents a maximum pressure on the axis of symmetry/rotation, whereas the total pressure for model M3 decreases on the axis. Further investigation is necessary to relate the spin of the black hole to the pressure decrease in the magnetic field cases. 

In models M1 and M2 we see that the magnetic pressure in the centre of the Mach cone has a much larger amplitude than the surrounding thermal pressure, while in model M3 the magnetic pressure has a smaller amplitude than the surrounding thermal pressure. Clearly, by allowing the magnetic field to ``wind'' with the black hole the magnetic pressure decreases which reduces the opening angle of the Mach cone. Since
we also see that when we turn on the rotation, that there is very little change in the opening angle between the hydrodynamic model and the magnetohydrodynamic model. However, we do need to perform a larger parameter survey to determine how the magnetic fields are affected by the rotation of the black hole.

\section{Accretion Rates\label{Sec:6}}
%+++++++++++++++++++++++++++++++++++++++++++++++++++++++++++++++++
\indent In this section, we discuss the measured quantities that we observe for each run. The first several are simply the same quantities as developed by Petrich \etal~\cite{PSST} and later used by Font \etal~\cite{FI1,FI3}. The accretion rates come from conservation laws at asymptotic infinity. We calculate the mass accretion rate from the mass of the accreted matter,
\begin{equation}\label{eq:mrate1}
m=\int{d^3x\rho}
\end{equation}
which when we take the time derivative and substitute the mass flux into the integral becomes,
\begin{equation}\label{eq:mrate2}
\dot{m}=\int{d^3x\partial_i(\rho \hat{v}^i)}=4\pi\oint{\sqrt{-g}\rho \hat{v}^r d\theta}.
\end{equation}
We calculate the energy and momentum accretion rates in a similar fashion,
\begin{equation}\label{eq:rates}
Q^{(\iota)}=-4\pi\int{d^3xT^{\mu\nu}n_{\mu}\preup{(\iota)}{\xi}_{\nu}}
\end{equation}
where $\preup{(\iota)}{\xi}_{\nu}$ is the $\iota$-th Killing vector of the system, and $n_{\mu}$ is a normal vector to the $3+1$ hypersurface.
With the given $3+1$ description, we use $n_{\mu}=(-\alpha,0,0,0)$ so the above equation reduces to,
\begin{equation}\label{eq:rates2}
Q^{(\iota)}=4\pi\int{d^3xT^{t\nu}\alpha \preup{(\iota)}{\xi}_{\nu}}.
\end{equation}
For the axisymmetric black hole there are two Killing vectors, one temporal and one azimuthal. These lead to conservation of energy and azimuthal angular momentum. At asymptotic infinity, the metric takes the form of Minkowski space which is maximally symmetric \cite{FI3}. Thus we may consider the radial Killing vector. By the same logic we are able to use the polar Killing vector as well; however, due to the symmetries involved in this study, no relevant physics can be extracted from this value.

To obtain the momentum accretion rates, we take the time derivative of Eqn.~\eqref{eq:rates2}, where the time derivative commutes with the spatial integral. Subsequently, we replace the time derivative quantities with their flux counterpart to get an expression for the accretion rates \cite{FI1,Donmez:2010}. Since our HRSC scheme is at most second order, we use a second order numerical integration scheme to solve the resulting expressions.

The quantities we are specifically interested in presenting are the radial momentum accretion rate, the energy accretion rate, and the mass accretion rate.

The first quantity considered is the energy accretion rate. The equation is stated here and is easily derived using Eqn.~(\ref{eq:rates}).
\begin{eqnarray}
Q^{(t)}=E&=&-\int{\sqrt{-g}\alpha T^{tt}d^3x} = -\int{\sqrt{\gamma}\alpha^2 T^{tt}d^3x}\nonumber\\
 &=& -\int{\sqrt{\gamma}E d^3x} \nonumber\\
\dot{Q}^{(t)}=\dot{E} &=& \oint{\sqrt{-g}T^{tr}d\theta} \nonumber\\&&+ \int_{r_{\rm{EH}}}^{r_m}{\sqrt{-g}T^{\mu\nu}\Gamma^t{}_{\mu\nu}drd\theta}.
\end{eqnarray}
Likewise we have the momentum accretion rates
\begin{eqnarray}
 \dot{Q}^{(r)}=\dot{P}^r&=&\oint{\sqrt{-g}T^{rr}d\theta}\nonumber\\&&+\int_{r_{\rm{EH}}}^{r_m}{\sqrt{-g}T^{\mu\nu}\Gamma^{r}{}_{\mu\nu} drd\theta},
\end{eqnarray}
where $r_{\rm{EH}}$ is the radial location of the black hole event horizon and $r_{m}$ denotes the radius that we measure the accretion rate. In our measurements, we are interested in capturing dynamics on the event horizon, so $r_m=r_{\rm{EH}}$, thus all volume integrals above are zero.

% The accretion process is an efficient method of producing large amounts of energy \cite{Shapiro}, considering that a body that falls directly into a Schwarzschild black hole will, according to an observer at infinity, slow to a halt and disappear as it penetrates the event horizon. In the event that the same matter orbits the black hole before it falls in, a portion of the rest energy of the particle will convert to heat and other forms of radiation. In the case of a rotating black hole, the time required to fall into the black hole is greater since the event horizon is located closer to the black hole; therefore, even more energy may be released before the particle is lost. The accretion luminosity, the measure of the radiation lost from the system, is proportional to the mass accretion rate $\dot{M}$ \cite{bob}. Petrich \etal~\cite{PSST} show that there is a linear relationship between the mass accretion rate and the energy accretion rate

To normalize the mass accretion rate, previous authors used the mass accretion rates determined at the sonic point for a relativistic Bondi (stationary) accretor \cite{PSST, FI1},
\begin{equation}
 \dot{M} = \frac{4\pi\lambda M^2\rho_{\infty}m_{B}}{(v_{\infty}^2+(c_{s}^{\infty})^2)^{3/2}}.
\end{equation}
$\lambda$ is a dimensionless form factor as described in Shapiro and Teukolsky \cite{Shapiro}. When normalizing the radial momentum accretion rate, they scaled the solution by both the mass accretion rate and the asymptotic velocity.
\begin{equation}
 \dot{P}^r = \frac{4\pi\lambda M^2\rho_{\infty}m_{B}v_\infty}{(v_{\infty}^2+(c_{s}^{\infty})^2)^{3/2}} = \dot{M}v_\infty.
\end{equation}
We do not use these factors for two reasons. First, we measure the accretion rates at the event horizon not the accretion radius; second, we are measuring the accretion rates for a magnetohydrodynamic system, and the above simple forms do not take into account magnetic effects. Instead we time evolve the relativistic Bondi--Hoyle system using $v_{\infty}=0$, therein solving the relativistic magnetized Bondi problem, starting from a uniform density background. When the flow reaches a steady state, we use those values of the accretion rates to normalize the accretion rate measurements for the relativistic Bondi--Hoyle systems.

% If no instabilities are present in the system, we expect the polar angular momentum accretion to remain at zero, or at worst oscillate within numerical tolerance around zero. Numerical roundoff errors account for the expected oscillations.

Due to the symmetry of the system the azimuthal velocity field is exactly zero and, consequently, is not plotted.
%antisymmetric about the $0-\pi$ axis. This leads the equation for the azimuthal accretion to tend to zero as it is antisymmetric over a symmetric domain. Deviations from this behaviour are observed below the lowest tolerance setting. Any strong deviation from this behaviour would have been an indication of an asymmetric flow forming.

% With the spherical body accreting matter, we expect all angular accretion to be zero by symmetry conditions as determined earlier in Font's papers \cite{FI1,FI2}. For smooth flows, small deviations from this result will be a numerical artifact. We monitor the growth of these quantities to determine when rounding error has accumulated sufficiently that the resulting solutions are no longer credible. Evidence has shown that the accumulation of the roundoff errors appear very late in the evolution, well past the time required to obtain a steady state solution.

% The final steady state solutions were found to be oscillatory about a fixed point and as resolution increased, these oscillations were found to decrease in amplitude. The final data presented is an average over the last 200 data points, the equivalent of averaging over that last $30M$ time steps.
We see that in all accretion plots that the flows reach a steady state. We also see the general trend that all magnetic flows have a wider Mach cone opening angle and consequently all magnetic flows experience a larger mass and energy accretion rate. This trend also extends to the radial momentum accretion rate for all models with $\Gamma=4/3$; however, for models with $\Gamma=5/3$ we find that this accretion rate decreases for larger amplitudes of the magnetic field. This is attributed to the relative build up of material in the downstream region. For all models studied we have an evacuated region downstream; however, it is observed that the models with $\Gamma=4/3$ have a narrower high magnetic pressure region, which leads to a narrower region of the shock tail that experiences density evacuation. As such this leaves a larger region of the Mach cone in contact with the black hole, and clearly, overall larger accretion rates. For $\Gamma=5/3$ models the magnetic pressure causes a much larger density depletion region. The wider magnetic pressure profile decreases the contact between the matter and the black hole, decreasing the drag between the two. The magnitude of the density in contact with the black hole is larger than the purely hydrodynamic simulation, accounting for the larger mass and energy accretion rates. This is also interesting behaviour, since it indicates that a cold fluid will experience less drag than the hot fluid.

\begin{figure}
 \includegraphics[width=3in]{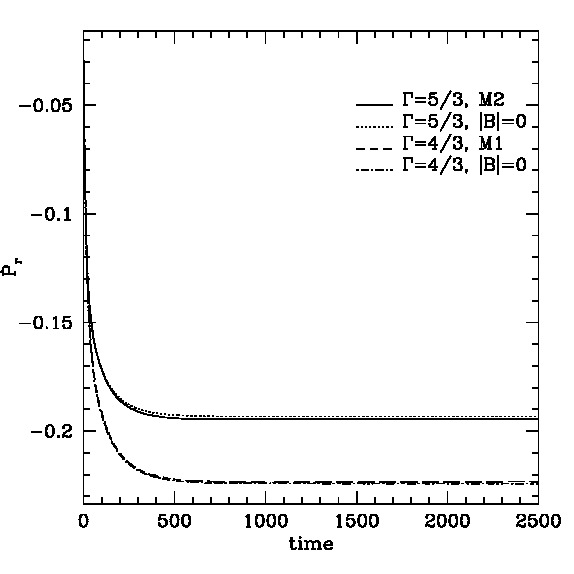}\\
 \caption{The radial momentum accretion rates for models M1, M2. We see that the adiabatic constant plays a significant role in the amount of drag the moving black hole experiences, which agrees entirely with previous studies.}\label{fig:drag_M1_M2}
\end{figure}

\begin{figure}
 \includegraphics[width=3in]{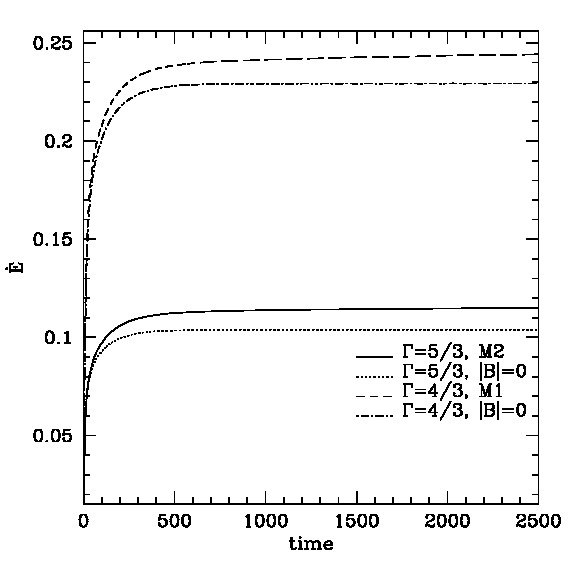}\\
 \caption{The energy accretion rates for models M1, M2. When comparing the Mach cone opening angles between models M1 and M2, we see that the accretion rates are related to the opening angle, the larger the cross section in the shock cone, the more energy in accreted. This finding is in agreement with the results of the purely hydrodynamic models in \cite{FI1}.}\label{fig:ener_M1_M2}
\end{figure}

\begin{figure}
 \includegraphics[width=3in]{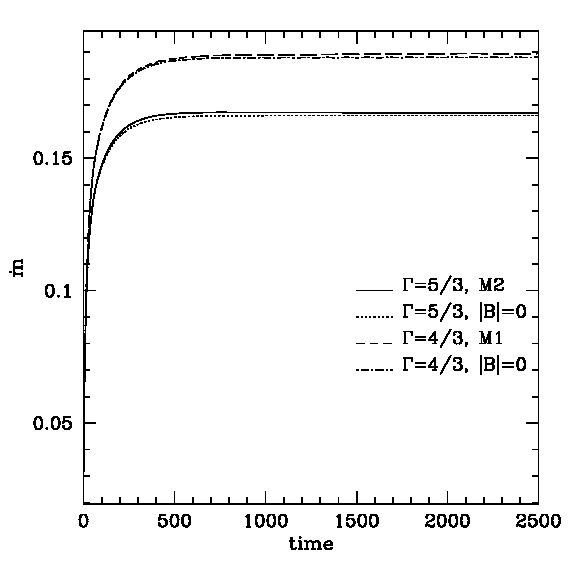}\\
 \caption{The mass accretion rates for models M1, M2. Just as in the case of drag, the mass accretion rate is significantly affected by the value of the adiabatic constant used.}\label{fig:mass_M1_M2_G43}
\end{figure}

In Figs.~\ref{fig:drag_M1_M2}, \ref{fig:ener_M1_M2}, and \ref{fig:mass_M1_M2_G43} we compare the accretion rates for the radial momentum, the energy and the mass. Just as presented in \cite{FI1}, the accretion rates are dependent on the choice of adiabatic constant. 

\begin{figure}
 \includegraphics[width=3in]{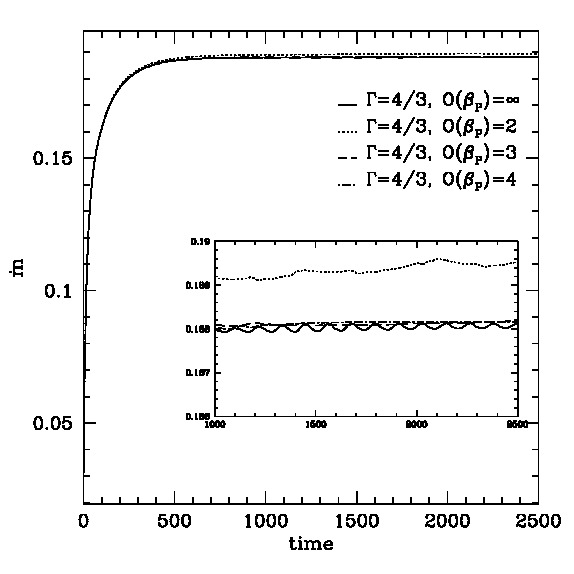}
 \caption{The mass accretion rates for models M1, M4, M5, and H1. We see that the mass accretion rate is only significantly impacted by the presence of a large magnetic field as simulated in model M1. As we expect, the mass accretion rates decrease as a function of the magnitude of the plasma beta parameter. What is of particular interest is the oscillatory behaviour of the purely hydrodynamic model, investigated by \cite{Donmez:2010} is damped out quickly by the presence of even a weak magnetic field as presented in model M5.}\label{fig:mass_M1_H1}
\end{figure}

\begin{figure}
 \includegraphics[width=3in]{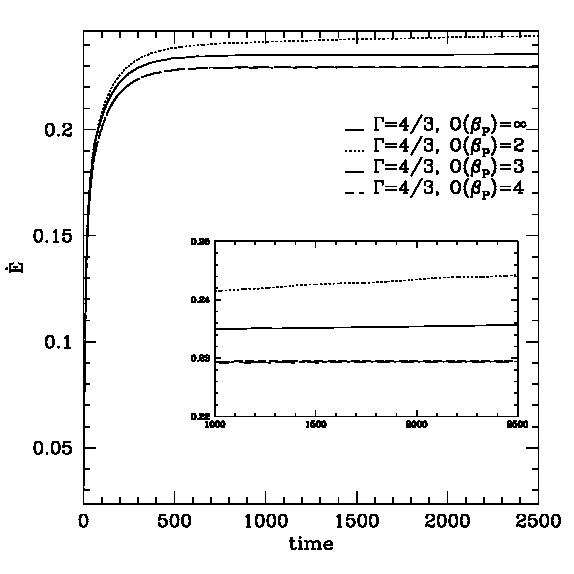}
 \caption{The energy accretion rates for models M1, M4, M5, and H1. Just as in Fig.~\ref{fig:mass_M1_H1}, the energy accretion rate is only significantly impacted by the presence of a large magnetic field. Indicating that the increase in the Mach cone opening angle allows more fluid to accrete.}\label{fig:ener_M1_H1_G43}
\end{figure}

\begin{figure}
 \includegraphics[width=3in]{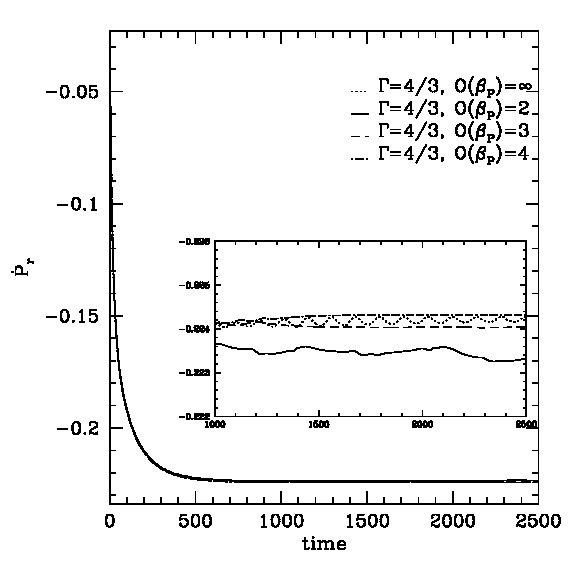}
 \caption{The radial momentum accretion rates for models M1, M4, M5, and H1. We see that like all the previous measurements for these models that the larger opening angle leads to an increased drag experienced by the star.}\label{fig:drag_M1_H1_G43}
\end{figure}

Figures \ref{fig:mass_M1_H1}, \ref{fig:ener_M1_H1_G43}, \ref{fig:drag_M1_H1_G43} display the mass, energy and radial momentum accretion rates for models M1, M4, M5, and H1. As we expect, as the magnetic field reduces in magnitude, the accretion rates approach the hydrodynamic rates. In the blown up region within each plot we see the accretion rate as a function of the plasma beta parameter. The energy and mass accretion rates decrease as a function of the magnetic field strength, and converge to the hydrodynamic accretion rate in the zero magnetic field limit. Although we do note that in Fig.~\ref{fig:mass_M1_H1}, the presence of even a weak magnetic field damps out the oscillatory accretion rate present in the hydrodynamic study. The oscillatory behaviour of the hydrodynamic solution is discussed in \cite{FI1, Donmez:2010}. The radial momentum accretion rates also increase as a function of the magnetic field strength.

%
% Now the gamma=5/3 plots
%

\begin{figure}
 \includegraphics[width=3in]{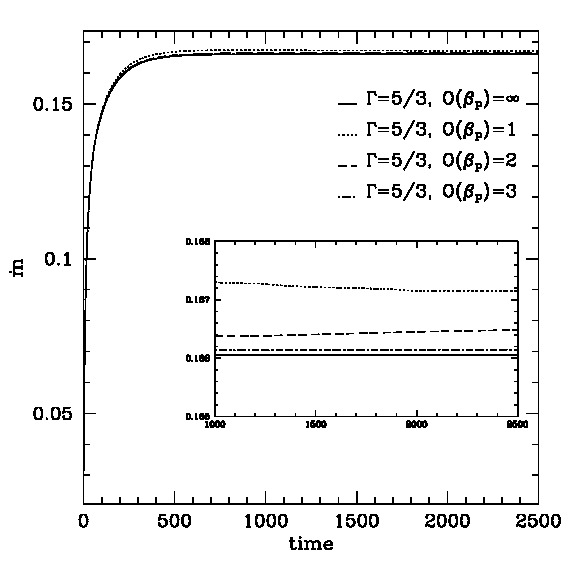}\\
 \caption{The mass accretion rates for models M2, M6, M7, M8, and H2. We see that the mass accretion rate is only significantly impacted by the presence of a large magnetic field as simulated in model M2. As we expect, the mass accretion rates decrease as a function of the magnitude of the plasma beta parameter. What is of particular interest is the oscillatory behaviour of the purely hydrodynamic model, investigated by \cite{Donmez:2010} is damped out quickly by the presence of even a weak magnetic field as presented in model M5. We note that for this adiabatic constant the oscillations in the hydrodynamic simulation do not exist. The mass accretion rate increases monotonically as a function of the plasma beta parameter.}\label{fig:mass_M2_H2}
\end{figure}

\begin{figure}
 \includegraphics[width=3in]{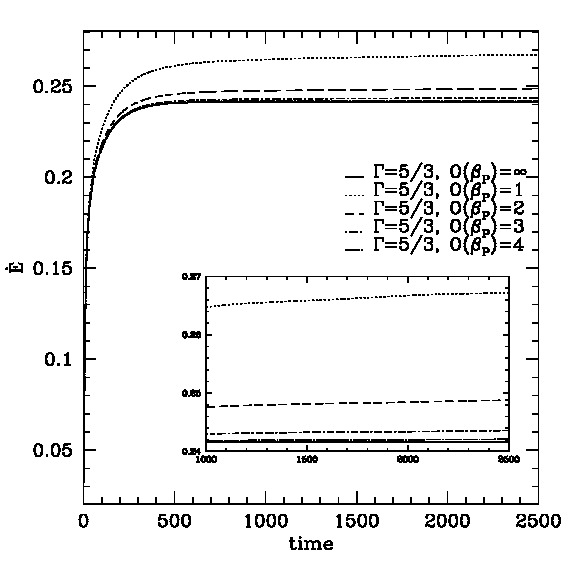}\\
 \caption{The energy accretion rates for models M2, M6, M7, M8, and H2. Just as in Fig.~\ref{fig:mass_M2_H2}, the energy accretion rate is only significantly impacted by the presence of a large magnetic field, indicating that the increase in the Mach cone opening angle allows more fluid to accrete. As is expected, the energy accretion rate increases in a similar fashion as the mass accretion rate \cite{PSST}.}\label{fig:ener_M2_H2}
\end{figure}

\begin{figure}
 \includegraphics[width=3in]{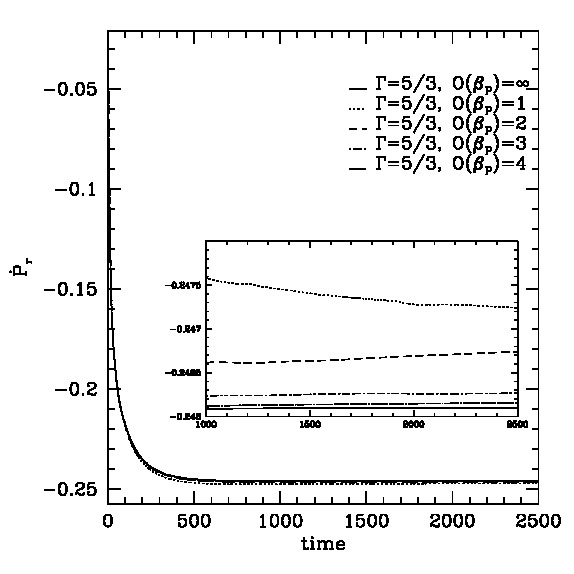}\\
 \caption{The radial momentum accretion rates for models M2, M6, M7, M8, and H2. We see that unlike all the previous measurements for these models that the larger opening angle leads to an decreased drag experienced by the star.}\label{fig:drag_M2_H2}
\end{figure}

Figures \ref{fig:mass_M2_H2}, \ref{fig:ener_M2_H2}, \ref{fig:drag_M2_H2} present the mass, energy and radial momentum accretion rates for models M2, M6, M7, M8, and H2. As we expect from previously discussed models, as the magnetic field reduces in magnitude, the accretion rates approach the hydrodynamic rates. In this case we find that the magnitude of the radial momentum accretion actually decreases with increased magnetic field strength. This is in contrast to the hot plasma case, where the drag experienced by the spherically symmetric black hole increases with increased magnetic field strength.

\begin{figure}
 \includegraphics[width=3in]{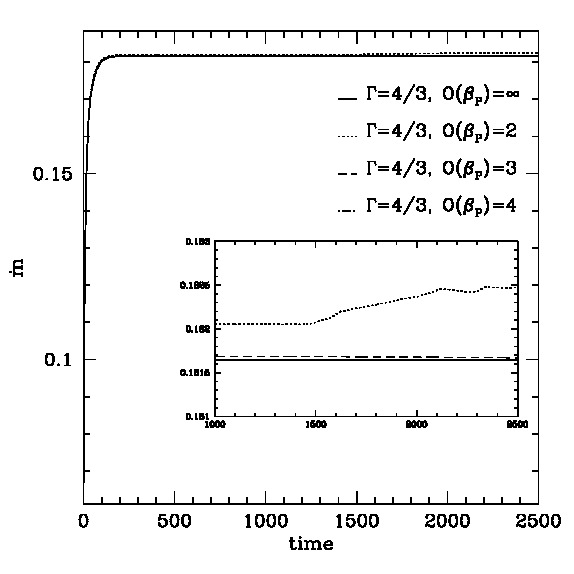}\\
 \caption{The mass accretion rates for models M3, M9, M10, and H3. We see that the mass accretion rate is only significantly impacted by the presence of a large magnetic field as simulated in model M3. As we expect, the mass accretion rates decrease as a function of the magnitude of the plasma beta parameter. First, we see no evidence of oscillatory mass accretion as seen in model H1. We note that for this adiabatic constant the oscillations in the hydrodynamic simulation do not exist. The mass accretion rate increases monotonically as a function of the plasma beta parameter. The mass accretion rate quickly diverges towards the end of the simulation. This is determined to be a result of the resolution of the simulation.}\label{fig:mass_M3_H3}
\end{figure}

\begin{figure}
 \includegraphics[width=3in]{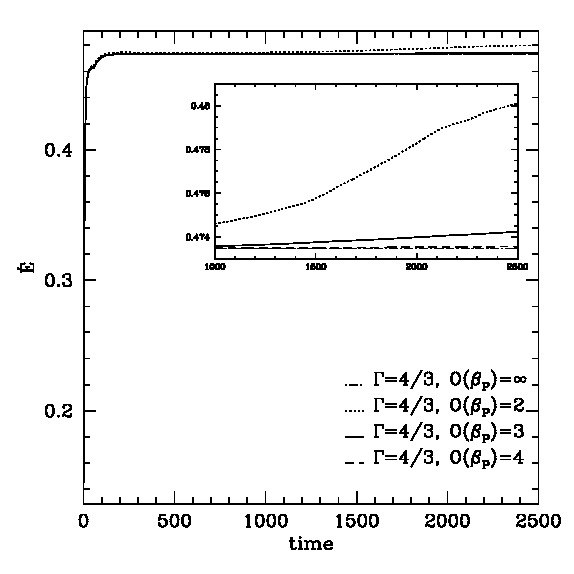}\\
 \caption{The energy accretion rates for models M3, M9, M10, and H3. Just as in Fig.~\ref{fig:mass_M3_H3}, the energy accretion rate is only significantly impacted by the presence of a large magnetic field, indicating that the increase in the Mach cone opening angle allows more fluid to accrete. As is expected, the energy accretion rate increases in a similar fashion as the mass accretion rate \cite{PSST}.}\label{fig:ener_M3_H3}
\end{figure}

\begin{figure}
 \includegraphics[width=3in]{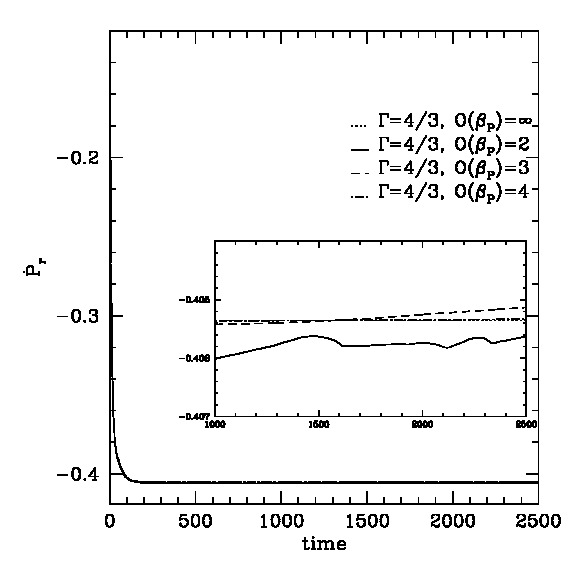}\\
 \caption{The radial momentum accretion rates for models M3, M9, M10, and H3. We see that unlike all the previous measurements for these models that the larger opening angle leads to an decreased drag experienced by the star.}\label{fig:drag_M3_H3}
\end{figure}

%
% Now the A=0.5 gamma=4/3 plots
%
Figures \ref{fig:mass_M3_H3}, \ref{fig:ener_M3_H3}, \ref{fig:drag_M3_H3} show the mass, energy and radial momentum accretion rates for models M3, M9, M10, and H3. These plots indicate that when accreting onto an axisymmetric black hole the general trends of the flow do not alter significantly. Just as in Figs.~\ref{fig:mass_M1_H1}, \ref{fig:ener_M1_H1_G43}, and \ref{fig:drag_M1_H1_G43}, the presence of a magnetic field increase all accretion rates. 
% CANT DO THE COMPARISON DIFFERENT VELOCITIES IN SPIN AND NO SPIN CASES
%For a comparison of models M1 and M3 directly we refer the reader to Figs.~\ref{}.
%\clearpage

The simulations presented here all had the property that they converge to the asymptotic solution shortly after $600M$.

\section{Flow Morphology\label{Sec:4}}
%+++++++++++++++++++++++++++++++++++++++++++++++++++++++++++++++++
The results of the simulations using the parameters found in table \ref{table:1} are discussed here. We present the final state of a sampling of the parameters surveyed in the spherically symmetric evolution. We see that the thermodynamic quantities establish a steady state solution. As in the general hydrodynamic models, we find that in the upstream region the contours are smooth, but in the downstream region there is the presence of a Mach cone. As with the original hydrodynamic case we also look at the profile along the $\phi=0$ and $\phi=\pi$ lines.

The simulations result in a rest mass evacuation immediately downstream of the black hole. Upon further analysis, this is the result of a buildup of the magnetic pressure downstream in the same region. The baryonic particles are transported along the magnetic field lines away from this region. As we see in figures \ref{fig:all_P_M1} and \ref{fig:all_P_M2}, the convergence tests as seen in Figs.~\ref{fig:M1converge} and \ref{fig:M3converge2} indicate that the region is tending to zero rest mass density. This phenomenon is similar to the behaviour of the particles in the solar wind as they interact with the earth's magnetopause \cite{Wang:2004}. In that setup, \cite{Wang:2004}, a region known as a plasma depletion layer develops in the upstream side of the earth moving through the solar wind from the sun. In said region the magnetic field strength increases and as a result the plasma density decreases. Although this process is most dominant in the upstream side of the earth's trajectory there is evidence that it also exists for the downstream side when the magnetic reconnection is insufficient to relieve the pile-up of magnetic field lines. In our system, the pile-up occurs primarily in the downstream side, which is likely due to the fact that the fluid itself is magnetic, and that the black hole is not assumed to have its own magnetosphere in our model. Future models may include such features, but such a study would be purely academic, since uncharged black holes are not expected to have their own magnetic fields.

Following the flow from upstream to downstream within a few Schwarzschild radii of the black hole reveals a very similar morphology as the hydrodynamic models. The flow is drawn to the black hole via gravitational attraction. As the material flows past the black hole, it is attracted to the hole by gravitational forces. The angular momentum of the fluid prevents the direct inflow of all the gravitationally bound material. This angular momentum is lost as the material moves downstream and begins to converge on the axis of symmetry. Here the fluid increases in density and pressure. This is where the similarities end. Closer to the black hole, we see a noticeable difference. While the same flow process occurs the effects of the magnetic field are far more obvious. The downstream pressure increases, including the magnetic pressure. As more fluid accumulates the frozen flux tubes also accumulate, which increases the local magnetic field strength. Eventually the magnetic pressure begins to dominate the local flow, and the density in that region begins to decrease as the fluid pressure outside of this region deflects new fluid from entering, but at the same time the magnetic pressure decelerates the fluid and directs it into the black hole. A balance between magnetic pressure and thermodynamic pressure is established and the flow reaches a steady state.

% The motion of the plasma is controlled by two forces, the pressure gradient of the fluid, and the force from the magnetic field. 
% The dominant force determines the behaviour of the different regions. ??
In Fig.~\ref{fig:cross_P_M1} we have a cross section of model M1. We see that the total pressure of the system appears continuous; however, when we look at the contributions from the different pressures the story is very different. In the evacuated region, we see that the magnetic pressure is dominant, showing that the force due to the magnetic field is strongest in this region. The radial velocity of the fluid in that region becomes negative indicating flow into the black hole, which is being forced faster than a hydrodynamic model due to the pressure from the magnetic field. As the oncoming flow reaches the tail shock, it first interacts with a region dominated by the thermodynamic pressure. If the fluid velocity is great enough the fluid will pass the first shock point, along the way interacting with a relatively minor build-up of magnetic field within the thermal pressure wall. Once past the wall, the fluid reaches a region where the thermal pressure suddenly drops and the magnetic pressure dramatically increases. It is in this region that the magnetic field is at its strongest, which forces the fluid to follow the path of least resistance, either the fluid accretes onto the black hole or flowing faster upstream, either way causing a density depletion. The low point in the magnetic pressure denotes the location where the magnetic field changes direction and points towards the black hole. This is seen most easily in Fig.~\ref{fig:mag_cont}.

\begin{figure*}[htp]
\begin{center}
  \includegraphics[width=2in, height=2in]{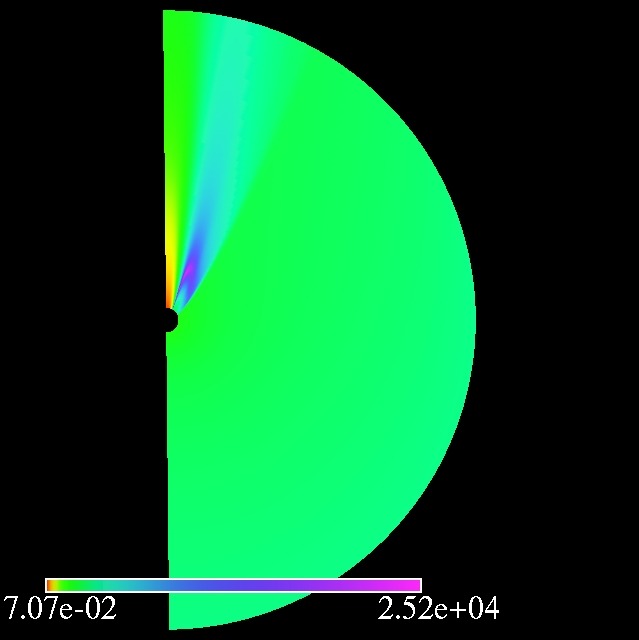}
  \includegraphics[width=2in, height=2in]{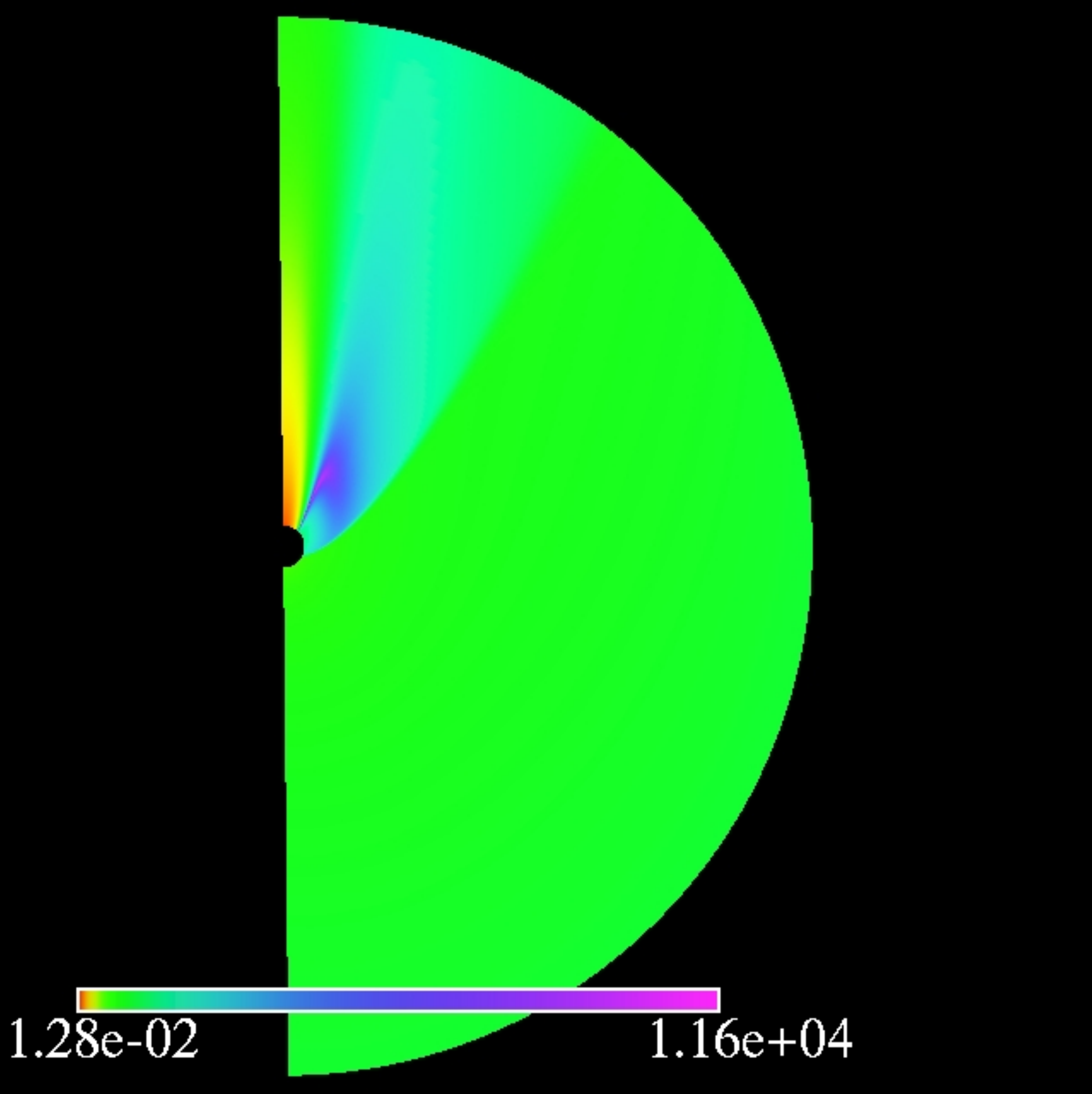}
  \includegraphics[width=2in, height=2in]{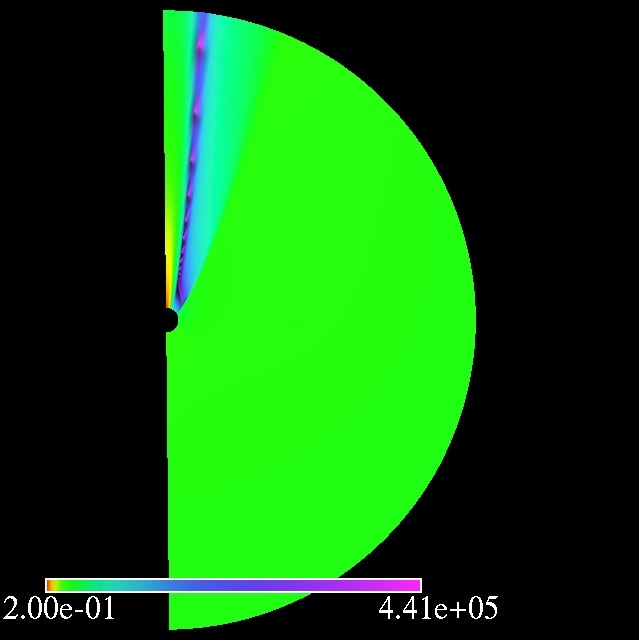}
\end{center}
\caption{Plasma beta parameter, $\beta_P(r,\theta)$ for the axisymmetric evolution for model M1 (left, $t=2500M$), M2 (centre, $t=1500M$), and M3 (right, $t=2500M$). These snapshots were taken well after the flow reaches its steady state. We see that a black hole travelling well above supersonic speeds produces a readily apparent tail shock. Also note the peaks that form along the inside edge of the tail shock for model M3. This is a feature only seen in simulations with a rotating black hole. The plasma beta parameter in the tail clearly follows a similar form as the tail shock near the black hole; however, for the spherically symmetric black holes, as we move further away downstream from the black hole the plasma beta parameter begins to smoothly reach asymptotic values. The rotation of the black hole clearly winds the magnetic field which may be seen in the maximum of $\beta_P$ in model M3 being an order of magnitude larger than that seen in model M1.}\label{fig:plasma_beta}
\end{figure*}

When investigating axisymmetric Bondi--Hoyle accretion onto a rotating black hole we found that the general flow morphology was substantially different to the non-rotating black hole with the same equation of state. Which is most dramatic when observing the plasma beta parameter as a function of the computational domain in Fig.~\ref{fig:plasma_beta}.

%%%%%%%%%%%%%%%%%%%%%%%%%%%%%%%%%%%%%%%%%%%%%%%%%%%%%%%%%%%%
% Insert Axisymmetric figures
%%%%%%%%%%%%%%%%%%%%%%%%%%%%%%%%%%%%%%%%%%%%%%%%%%%%%%%%%%%%

Finally, in Fig.~\ref{fig:plasma_beta} we present the plasma beta parameter through the domain, after each simulation has reached a steady state. We see that a black hole travelling well above supersonic speeds produces a readily apparent tail shock. Also note the peaks that form along the inside edge of the tail shock for model M3. This is a feature only seen in simulations with a rotating black hole. The plasma beta parameter in the tail clearly follows a similar form as the tail shock near the black hole; however, for the spherically symmetric black holes, as we move further away downstream from the black hole the plasma beta parameter begins to smoothly reach asymptotic values. The rotation of the black hole clearly winds the magnetic field which may be seen in the maximum of $\beta_P$ in model M3 being an order of magnitude larger than that seen in model M1.

%

%
%+++++++++++++++++++++++++++++++++++++++++++++++++++++++++++++++++
\section{Conclusions\label{Sec:7}}
%+++++++++++++++++++++++++++++++++++++++++++++++++++++++++++++++++
In this paper we presented a preliminary survey of the magnetized Bondi--Hoyle accretion onto both spherically symmetric and rapidly rotating axisymmetric black holes with the rotation axis aligned with the asymptotic uniform magnetic field. We used a high-resolution shock-capturing scheme with an approximate Riemann solver. To maintain the $\nabla\cdot B=0$ constraint, we used a hyperbolic-divergence cleaning method and used the auxiliary field as more than just a means to remove divergence violations, but also as a well behaved function that may be used to determine the convergence rates for magnetized systems.

We have established that when magnetic fields are introduced to the typical hydrodynamic calculations, the evolution of an axisymmetric system exhibits behaviour that is qualitatively broadly similar to the purely hydrodynamic counterpart 
in \cite{FI1,FI2} but differs quantitatively in a range of crucial ways.
%exhibits behaviour that is very different from the purely hydrodynamic counterpart investigated in \cite{FI1,FI2}. 
%
Ultimately, we have shown that all parameters surveyed resulted in a steady state solution with an evacuated region downstream of the black hole. This evacuation is a result of a build-up of magnetic pressure in the same region, which forces all matter out of this region. Future studies will investigate a wider range of parameters and will attempt to match the parameters investigated to astrophysical observations. We have also seen evidence that although a rotating black hole in a hydrodynamic simulation aligned in a configuration similar to ours yields very little in the way of different morphology, the introduction of an ideal magnetic field is sufficiently different to make the dynamics much more interesting.

The presence of the magnetic field is seen to increase the Mach cone opening angle, and consequently increase the accretion rates for both the mass and energy. However, we see that the radial momentum accretion rates, or drag, decreases as a function of the magnetic field strength for cold plasmas, while hot plasmas experience more drag as a function of the magnetic field strength. This is attributed to the fact that the magnetic pressure of a $\Gamma=5/3$ fluid downstream of the black hole is wider (relative to the $\Gamma=4/3$ models), and creates a wider evacuated region, and less surface area of the black hole event horizon is exposed to the flow in a region where all accretion is expected to be maximal.

% The presence of the magnetic field had only a small impact on the mass and radial momentum accretion rates. This could be due to two factors; one, that the magnetic fields investigated in this study are too small to have an impact, which leads us to believe that more advanced numerical methods will be necessary and incorporating full electrodynamic hydrodynamic models such as those proposed by Palenzuela \etal~\cite{Palenzuela:2008}. The second conclusion from the lack of impact is that the magnetic pressure inside the evacuated region actually pushes the fluid further downstream of the black hole, which allows it to escape the gravitational attraction of the black hole. To discern between these possibilities, we will need further investigation.

This was a phenomenological study, a detailed analysis will require the use of less diffusive Riemann approximations such as the Roe or Marquina solvers as used in recent hydrodynamic studies of the Bondi--Hoyle accretion problem in \cite{Donmez:2010}. Further, a detailed study of the parameters needed in the hyperbolic diffusion cleaning method will need to be refined. Future studies will also involve a full three dimensional code which will allow a general asymptotic magnetic field configuration as prescribed by Bi\v{c}\'{a}k \cite{Bicak:1985}, in order to capture more astrophysically justifiable configurations. This line of study is far from complete, as this configuration will provide an excellent and simple testbed for more general fluid models as the field grows increasingly more complicated. 

\begin{acknowledgments}
The author would like to thank the scientific grant agencies NSERC and CIFAR for funding this project, and acknowledge the contributions from the Relativity group at the University of British Columbia, as well as the members of the General Relativity group in the School of Mathematics at the University of Southampton who offered a lot of insightful comments and suggestions. In particular, the author would like to thank Ian Hawke for proof reading this article.
The simulations presented in this article were performed on computational resources supported by the PICSciE-OIT High Performance Computing Center and Visualization Laboratory.
This research has also been enabled by the use of computing resources provided by WestGrid and Compute/Calcul Canada.
\end{acknowledgments}

%+++++++++++++++++++++++++++++++++++++++++++++++++++++++++++++++++
\section{Appendix: Code Verification}\label{SubSec:1}
\begin{figure}
  \includegraphics[width=3in]{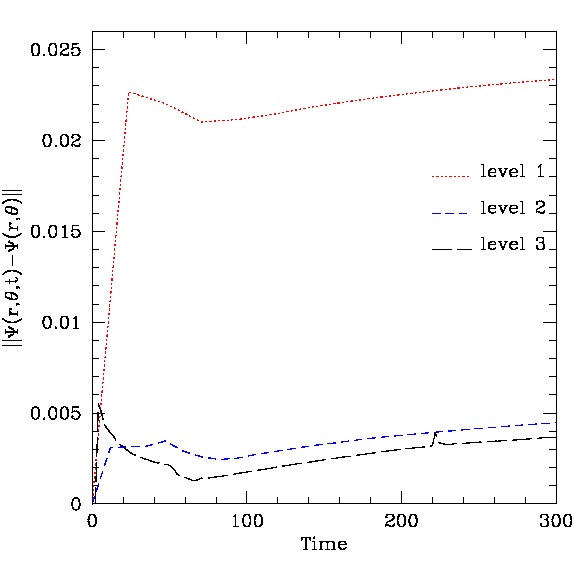}
\caption{The convergence of the $L_2$ norm of the auxiliary field $\psi$ for model M3. It is clear that the system is convergent. Level 1 uses a grid resolution $200\times80$, level 2 denotes $400\times160$, and level 3 denotes $800\times320$.}\label{fig:M3converge}
\end{figure}

\begin{figure}
  \includegraphics[width=3in]{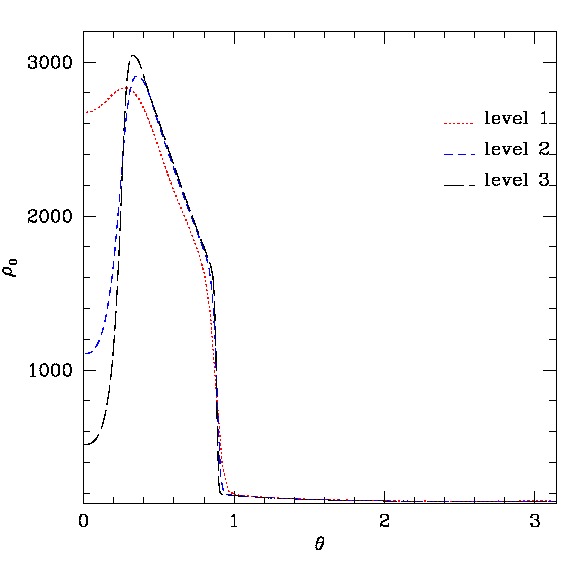}
\caption{The convergence of the rest mass density density $\rho_0$ cross section at $r=2M$ for model M1. We see that the evacuated region near the axis of symmetry is converging to zero. Level 1 uses a grid resolution $200\times80$, level 2 denotes $400\times160$, and level 3 denotes $800\times320$.}\label{fig:M1converge}
\end{figure}

\begin{figure}
  \includegraphics[width=3in]{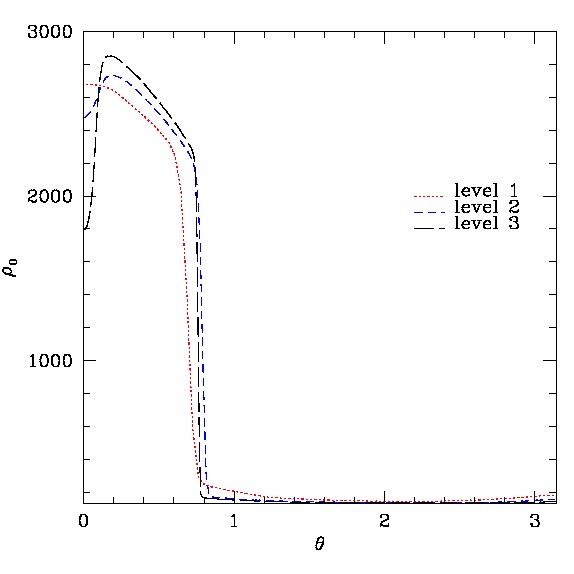}
\caption{The convergence of the rest mass density density $\rho_0$ cross section at $r=1.86M$ for model M3. We see that the evacuated region near the axis of symmetry is converging to zero. Level 1 uses a grid resolution $200\times80$, level 2 denotes $400\times160$, and level 3 denotes $800\times320$.}\label{fig:M3converge2}
\end{figure}

%+++++++++++++++++++++++++++++++++++++++++++++++++++++++++++++++++
To verify the numerical accuracy of the code we performed convergence testing of two different quantities. First we used a global quantity, the $L_2$-norm of the auxiliary variable $\Psi$, and second we used a cross section of the pressure accretion profile. We considered a cross section of the data at a constant radial coordinate, and independently a slice through a constant value of the angular coordinate. Different techniques for convergence testing HRSC schemes is an active research avenue and there is no universally accepted scheme at the time of writing. Convergence tests for models M1 and M3 may be seen in Figs.~\ref{fig:M3converge}, \ref{fig:M1converge} and \ref{fig:M3converge2}. While we see that the depletion region has not converged by our selected resolution, the various accretion rates are within convergence. Future work will investigate the different parameters using higher resolution to work in the convergent regime of the depletion region.

To determine the convergence of the divergence cleaning variable $\psi$ to zero, we calculate the $L_2$-norm of this scalar field. While this would at best indicate a first order convergence rate, the fact that this term converges to zero is critical in showing the validity of our selected method for handling divergence violations. A sample convergence test for the $\psi$ variable is seen in Fig.~\ref{fig:M3converge}.

The disagreement between the true second order convergence value and our system comes from the special treatment of shock regions and local maxima where the system reduces to first order convergence as mentioned above.

% \begin{figure}
%   \includegraphics[width=3in]{thesis_pics/axi_v_6_P_converge.pdf}
% \caption{A convergence test for model U2. We used a cross-section of the pressure variable in the upstream region (denoted by the negative radial direction) along the axis of symmetry. We see that the system is converging on a shock front upstream of the travelling black hole. Level 1 is $200\times80$, level 2 denotes $400\times160$, and level 3 denotes $800\times320$. In this convergence test we focused on the region $r_{\min}\le r\le 20$ to emphasize the shock front.}\label{fig:v6converge}
% \end{figure}

% \begin{figure}
% \begin{center}
% %\includegraphics[width=3in]{faster/fast_convergence.pdf}\\
% \includegraphics[width=3in]{thesis_pics/axi_v_9_P_converge.pdf}\\
% \caption{A convergence test for model U4. In this test we used a cross section along $r=4$ around the black hole. In this test we see that the solution is converging on a tail shock opening angle of approximately 1.4 $rad$, and a maximum downstream pressure of approx.~107. Level 1 is $200\times80$, level 2 denotes $400\times160$, and level 3 denotes $800\times320$.}\label{fig:v9converge}
% %\end{minipage}
% \end{center}
% %\caption{The radial momentum accretion rates convergence test for model U3 (top) and U9 (bottom). We see here that the final result converges as approximately second order.}\label{fig:Fconverge}
% \end{figure}

\bibliographystyle{plain}
\bibliography{mybib}

%%%%%%%%%%%%%%%%%%%%%%%%%%%%%%%%%%%%%%%%%%%%%%%%%%%
\end{document}